\documentclass[a4paper,11pt]{article}
\usepackage{jcappub}
\usepackage[T1]{fontenc}
\usepackage{graphics}
\usepackage{graphicx}
\usepackage{pstricks}
\newcommand{\be}{\begin{eqnarray}}
\newcommand{\beq}{\begin{equation}}
\newcommand{\eeq}{\end{equation}}
\newcommand{\ee}{\end{eqnarray}}

\newcommand{\bmp}{\noindent\begin{minipage}{16cm}}
\newcommand{\emp}{\end{minipage}\vskip 7mm} 

\usepackage{bm}
\usepackage{amsmath}
\usepackage{amsfonts}
\usepackage{bbm}
\newcommand{\erf}{\mathrm{erf}}
\def\lsim{\mathrel{\rlap{\lower4pt\hbox{\hskip1pt$\sim$}}
    \raise1pt\hbox{$<$}}}                
\def\gsim{\mathrel{\rlap{\lower4pt\hbox{\hskip1pt$\sim$}}
    \raise1pt\hbox{$>$}}}                
\title{On the  Capture of Dark Matter by  Neutron Stars} 

\author[a]{Tolga  G\"uver,}   
\author[b]{Arif   Emre   Erkoca,}   
\author[c]{Mary Hall Reno}
\author[b,d]{and Ina Sarcevic}

\affiliation[a]{Department of Astronomy and Space Sciences, Faculty of
  Sciences, Istanbul University, 34119 University, Istanbul, Turkey}

\affiliation[b]{Department of Physics,  University of Arizona, Tucson,
  AZ 85704}

\affiliation[c]{Department  of Physics  and  Astronomy, University  of
  Iowa, Iowa City, IA 52242}

\affiliation[d]{  Department  of  Astronomy and  Steward  Observatory,
  University of Arizona, Tucson, AZ 85721}
  
\emailAdd{tolga.guver@istanbul.edu.tr}
\emailAdd{aeerkoca@gmail.com}
\emailAdd{mary-hall-reno@uiowa.edu}
\emailAdd{ina@physics.arizona.edu}

\abstract{We  calculate the  number of  dark matter  particles that  a
  neutron star accumulates over its  lifetime as it rotates around the
  center  of   a  galaxy,   when  the  dark   matter  particle   is  a
  self-interacting boson  but does not self-annihilate.   We take into
  account dark matter  interactions with baryonic matter  and the time
  evolution  of the  dark matter  sphere  as it  collapses within  the
  neutron star.   We show that  dark matter self-interactions  play an
  important role in the rapid accumulation  of dark matter in the core
  of the neutron star.  We  consider the possibility of determining an 
  exclusion region  of the  
   parameter space for  dark matter  mass and
  dark matter interaction cross section with the nucleons as well 
  as dark matter self-interaction cross section, based on the observation of
  old  neutron stars.  We
  show that for  a dark matter density of $~10^3$  GeV/cm$^3$ and dark
  matter mass  $m_\chi\lsim 10$  GeV, there  is a  potential exclusion
  region  for dark  matter interactions  with nucleons  that is  three
  orders of  magnitude more stringent than  without self-interactions.
  { The potential exclusion region for dark matter 
  self-interaction cross sections is  many  orders  of
  magnitude stronger than the current Bullet Cluster limit.  For 
  example, for high dark matter density regions, we find that for 
  $m_\chi \sim 10$ GeV when the dark matter interaction cross 
  section with the nucleons ranges from 
  $\sigma_{\chi n} \sim 10^{-52}$ cm$^{2}$ to 
  $\sigma_{\chi n} \sim 10^{-57}$ cm$^{2}$, the dark matter self-interaction 
  cross section limit is $\sigma_{\chi \chi}  \lsim  10^{-33}$ cm$^{2}$, which is 
  about ten orders of magnitude stronger than the Bullet Cluster limit.  
}}

\begin{document}
\maketitle
\flushbottom

\section{Introduction}

There is overwhelming evidence for the existence of dark matter in the
Universe,  from the  observation of  missing mass  in galaxy  clusters
\cite{Zwicky1933}  to the  precise  measurements  of the  cosmological
baryonic  fraction performed  by WMAP\cite{WMAP}  and BOSS\cite{BOSS}.
The possibility that the standard gravitation law needs to be modified
to explain the observations with  the ordinary visible baryonic matter
has   recently   been  ruled   out   by   the  Bullet   Cluster   data
\cite{Zaritsky}.  The  particle physics interpretation of  dark matter
requires dark matter particles to be weakly interacting and in thermal
equilibrium until  the Universe expansion becomes  such that particles
cannot find each other and their interactions freeze-out.  Large-scale
structure formation  indicates that dark  matter particles need  to be
non-relativistic at the time of freeze-out, i.e., dark matter needs to
be  ``cold."  Measurements  of  the matter  density  and its  baryonic
component imply  that the  dark matter  density contribution  is about
$25\%$ \cite{KolbTurner,DMreview1,DMreview2}.

Since the  dark matter density  is inversely proportional to  the dark
matter annihilation cross section  at freeze-out, the observed density
of dark matter in the Universe today constrains the annihilation cross
section in the early Universe, specifically at the time of freeze-out.
On dimensional grounds, a dark matter  particle with mass in the range
of  $100$ GeV  to  several  TeV with  weak  scale  couplings can  have
annihilation  cross   sections  of   the  order  of   $\langle  \sigma
v\rangle_{ann}=3\times  10^{-26}$ cm$^3$  s$^{-1}$ at  the freeze-out,
providing  a natural  explanation  for the  observed  density of  dark
matter today \cite{KolbTurner}.  There have been recent discussions of
the  possibility of  asymmetric  dark matter,  where  the dark  matter
particles   are   not   self-conjugate,   see   for   example,   Refs.
\cite{ADM1,ADM2,Cui:2011qe,McDermott:2011vn,Zentner:2011wx,
  Davoudiasl:2013pda,  Zurek:2013wia,   Laha:2013gva}  and  references
therein.  An initial particle-antiparticle asymmetry ultimately leaves
non-annihilating DM particles remaining in  the current epoch. This is
the case we consider in this paper.

As an astronomical  object in the Galaxy rotates around  the center in
its orbit,  it will sweep  through the  Galactic dark matter  halo and
eventually capture  some of the particles  on its way.  In  time, dark
matter  particles   that  are  captured   may  have  effects   on  the
observational properties of the astronomical object, which may then be
used  to constrain  the nature  of the  dark matter  \cite{press1985a,
  Zentner:2009ys,
  Kouvaris:2010kx,McCullough,Zentner:2011wx,goldman1989a,Bertone:2007ae,
Sandin:2008db,Ciarcelluti:2010ji,Capela:2013yf,Pani:2014rca,
  Kouvaris:2008zr,   Kouvaris:2011fk,Kouvaris:2011uq,McDermott:2011vn,
  deLavallaz:2010wp,Kouvaris:2011gb,Bramante,review,bell,
  kusenko,kouvaris_heavy,Kouvaris:2013awa, Bramante:2013nma}.  In that
respect, neutron stars  provide a natural laboratory  to constrain the
properties                of                dark                matter
\cite{goldman1989a,Bertone:2007ae,Sandin:2008db,Ciarcelluti:2010ji,Capela:2013yf,Pani:2014rca,Kouvaris:2008zr,Kouvaris:2011fk,
  Kouvaris:2011uq,McDermott:2011vn, deLavallaz:2010wp,Kouvaris:2011gb,
  Bramante,review,bell,kusenko,kouvaris_heavy}.    Even   though   the
surface  area of  a typical  neutron star  is much  smaller than  more
traditional  astronomical objects  like the  Sun, two  properties make
neutron  stars  very  efficient  in  capturing  Galactic  dark  matter
particles.  First, the immense baryonic  density inside a neutron star
provides a  natural location where there  it is very likely  that dark
matter particles  will interact and  lose energy.  Second,  because of
the strong  gravitational force,  it is also  almost impossible  for a
dark matter particle to escape from  a neutron star once it loses some
of its energy through interactions.

It may be only  a matter of time for a neutron  star to capture enough
number  of   dark  matter   particles  to  affect   its  observational
properties.  If the  dark matter particles are  annihilating, one such
effect can be seen in the cooling  of an old neutron star.  The energy
outcome of the annihilation process will  result in an increase of the
temperature  that  will remain  constant  and  discernible from  other
cooling processes  in time  (see, e.g.,  Ref. \cite{Kouvaris:2008zr}).
Calculations of the  annihilation effects on the cooling  of a neutron
star show that  the resulting effective temperature of  a neutron star
would   be    approximately   3000$-$10000~K   \cite{Kouvaris:2008zr},
depending  on the  local dark  matter density,  and the  mass and  the
radius  of  the neutron  star.   However,  the emission  of  blackbody
radiation at these temperature peaks at the UV to optical wavelengths,
where the  Galactic extinction hampers our  observational capabilities
to obtain precise measurements of  the surface temperatures of neutron
stars unless they are very close.

Even if the dark matter is not annihilating, under certain conditions,
the  capture process  may  still have  observable  effects.  For  some
values of  the local  dark matter  density, dark  matter mass  and its
interactions  with  nucleons and  amongst  themselves,  the number  of
particles {may  be enough for the  dark matter to be  relativistic and
  accumulate  to numbers  larger than  the Chandrasekhar  limit.  Once
  dark matter inside a neutron star that has reached the Chandrasekhar
  limit and reaches the self-gravitating limit, it may collapse into a
  black hole, which  could destroy the whole neutron star.   In such a
  case, even the  very existence of neutron stars at  certain ages can
  be used to  constrain the properties of dark matter.   This has been
  studied  in case  when dark  matter has  interactions with  baryonic
  matter only \cite{goldman1989a,Kouvaris:2008zr,McDermott:2011vn}.

Neutron  star constraints  on non-annihilating  dark matter  have also
been recently  studied including perturbative self-interactions  via a
$\lambda                      \phi^4$                      interaction
\cite{Kouvaris:2011gb,Bramante,review,bell}.   The   introduction   of
self-interactions modifies  the non-interaction  bosonic Chandrasekhar
limit, but this is model dependent \cite{review,csw,Mielke}.  We focus
here on the effect of  dark matter with {\it strong} self-interactions
in the accumulation of dark matter in neutron stars. For spin-0 bosons
with  a  $\lambda \phi^4$  interaction,  we  are considering  coupling
constants well beyond  the perturbative regime. Our  interest here are
very  small dark  matter-nucleon  cross sections,  yielding slow  dark
matter thermalization  in the neutron star  and large self-interaction
cross  sections  which  yield  a  phase  of  exponentially  increasing
$N_\chi$. This is a regime not covered in the recent literature.

In  the  absence  of  a  rigorous  Chandrasekhar  limit  for  strongly
interacting  bosons,  we  use   the  minimal  non-interaction  bosonic
Chandrasekhar limit, evaluating the  parameter space where $N_\chi$ is
large  enough  to  be  both   self-gravitating  and  larger  than  the
Chandrasekhar limit for bosons.  We discuss the several stages of dark
matter  accumulation:  first the  initial  capture,  then dark  matter
energy loss  by scattering with  nucleons and eventually the  onset of
self-capture, as a function of time.  We find that only for large dark
matter densities, $\rho_\chi \gsim  10^3$ GeV/cm$^3$ and $m_\chi$ less
than tens of GeV, is the  dark matter number $N_\chi$ at $t_{max}=10^9$ year
large  enough to  satisfy  both the  minimal  Chandrasekhar limit  for
bosons and the requirement that  the dark matter be relativistic, even
for strong self-interactions. Given  the relatively rare occurrence of
such high  local dark  matter densities, and  the likelihood  that any
model  with self-interactions  will  increase the  minimal limit,  the
existence of neutron stars does  not unambiguously constrain even very
strongly interacting asymmetric boson theories.}

We start the paper in Section 2 by reviewing the conditions necessary
for a  neutron star to collapse  as it captures dark  matter particles
for both  fermionic and  bosonic dark  matter.  We  discuss conditions
necessary for dark matter particles  to be relativistic as they become
captured by  the neutron star.   {We discuss the  accumulation regimes
  and  characteristic  times,  and   eventual  thermalization  of  the
  captured dark matter particles. } For bosonic dark matter, we review
conditions for self-gravitation and discuss the case when bosonic dark
matter forms a Bose-Einstein condensate.

In Section 3,  we outline our calculation for the  time evolution of
the dark matter particles as they  get captured by a neutron star. {In
  particular,  new here  is  an evaluation  of the  effect  of a  time
  dependent geometric limit for dark  matter self-capture in a neutron
  star.}  In Section  3 we summarize our inputs  and parameters used
in the  calculation and give a  simple example and rough  estimate for
the range of dark matter cross sections that would result in providing
dominant effects on the increase of the number of captured dark matter
particles as evolved  in time.  We discuss our results  in Section 4,
followed by conclusions in Section 5.
 
\section{Conditions for the collapse of neutron star}

\subsection{Chandrasekhar limit}

Our focus is on dark matter collapse  to a black hole within a neutron
star. As discussed below, thermalization  of the dark matter particles
is  an  important  feature   for  fermionic  dark  matter.  Subsequent
accumulation to the limit of  self-gravity is important if bosons have
an  impact  on  neutron  star  collapse. We  begin  by  reviewing  the
well-known results for neutron star collapse.

For  neutron  stars,  collapse  occurs  only  when  the  nucleons  are
relativistic.  The   relativistic  energy  of  the   neutron  star  is
approximately
\begin{equation}
  E \sim - \frac{3}{5}\frac{GN_{n}^2m_{n}^2}{R}+\left(\frac{9}{32 \pi^2}\right)^{1/3}N_n^{4/3}\frac{\hbar c}{R},
\label{energy_fermion}
\end{equation}
where $G$  is Newton's  constant and  $m_n$ is  the neutron  mass, and
$N_n$ is the number of neutrons.  When the gravitational energy of the
neutrons  equals the  energy due  to the  relativistic Fermi  momentum
($E=0$),  the number  of neutrons  is determined,  independent of  the
radius $R$  of the neutron  star.  Additional neutrons  cause collapse
into a black hole. The limit  on $N_n$ when $E=0$ is the Chandrasekhar
limit for neutron stars,
\begin{equation}
N^{Ch}_{n} \approx \left(\frac{1}{Gm_n^2}\right)^{3/2} \approx
2.2\times 10^{57} .
\label{chandra_neutrons}
\end{equation}
More generally, for fermions with mass $m_\chi$, the Chandrasekhar limit is   
\begin{equation}
N^{Ch}_{f} \approx \left(\frac{1}{Gm_\chi^2}\right)^{3/2} \approx
1.8 \times 10^{51} \left(\frac{\rm{100\ GeV}}{m_{\chi}}\right)^3\ ,
\label{chandra_ferm}
\end{equation}
where  the interactions  with  neutrons are  neglected.  An  essential
feature in this derivation is that the fermions are relativistic.  The
average energy of a relativistic fermion contained within a radius $r$ must satisfy
\begin{equation}
E_f\simeq \frac{N^{1/3}_f\hbar c}{r}> m_\chi c^2\ .
\label{relfermion}
\end{equation}
This translates  to a  requirement that the  number of  fermions $N_f$
should be larger than $N_f^{rel}$,
\begin{equation}
\label{eq:nfrel}
N_f\geq N_f^{rel}= 1.6\times 10^{65} \Biggl( \frac{m_\chi}{100\ {\rm GeV}}\Biggr)^3
\Biggl(\frac{r}{10.6\ {\rm km}}\Biggr)^3\ ,
\end{equation}
in order that the fermions are relativistic. The number of neutrons in
a neutron star  with a typical radius of $r=R=10.6$  km and neutron star
mass in terms of the solar mass $M_\odot$,  $M=1.44 M_\odot$, is such  that the neutrons are non-relativistic.
For dark  matter fermions to cause  the neutron star to  collapse, the
requirement  for relativistic  energies  may not  be satisfied  unless
$r\ll 10.6$ km or the dark matter mass is small compared to $m_n$. The
typical radius  containing most of  the dark matter reduces  when dark
matter is thermalized with neutrons in the neutron star, in which case
$r$  is the  thermalization radius $r_{th}$, much smaller  than the  neutron star
radius. We return to this below.

For bosonic dark matter, there is no Fermi pressure. The corresponding
Chandrasekhar{-like}   limit   for   bosons,  again   neglecting   the
gravitational energy  associated with neutrons  and self-interactions,
is \cite{McDermott:2011vn}
\begin{equation}
N^{Ch}_{b} \approx \frac{1}{Gm_\chi^2} \approx
1.5 \times 10^{34}\left(\frac{100\  {\rm GeV}}{m}\right)^{2}\ ,
\label{chandra_bos}
\end{equation}
since the kinetic energy per boson  is of order $E\sim \hbar c/r$, the
zero point energy  due to the uncertainty  principle. This approximate
limit is  confirmed by  a number  of approaches  to calculating  the {
  bosonic version  of the}  Chandrasekhar limit  \cite{ruffinib}.  The
introduction  of self-interactions  modifies  this  limit for  bosons,
potentially raising the limit to  close to the Chandrasekhar limit for
fermions   \cite{csw,Mielke},  although   this   is  model   dependent
\cite{review}.  {More recent  discussions  with perturbative  $\lambda
  \phi^4$ and  relatively quick dark matter  thermalization appear in,
  for  example,   Refs.  \cite{Kouvaris:2011gb,Bramante,review,bell}.}
Because  we are  considering  non-perturbative self-interaction  cross
sections, for this  paper, we rely on eq.  (\ref{chandra_bos}) for the
Chandrasekhar  limit for  bosons.  As  we see  below,  even with  this
relatively weak constraint on the number  of bosons required to form a
black hole, large ambient dark  matter densities are required for this
limit to be reached.

If the  dark matter  particles are bosons,  they are  not relativistic
unless they get  trapped inside the small region of  the neutron star.
The  requirement   for  relativistic  bosons  is   that  $E\sim  \hbar
c/r>m_\chi^2$, so the relativistic condition is
\begin{equation}
\label{eq:bosonrel}
\frac{m_\chi}{100\ {\rm GeV}}\frac {r}{10.6\ {\rm km}}< 2\times 10^{-22}\ .
\end{equation}  
After bosonic  dark matter  is thermalized, it  is primarily  within a
small   radius,   but   that   radius  is   not   small   enough   for
eq. (\ref{eq:bosonrel}) to be satisfied.  A larger dark matter density
is required  for the bosons to  form a black  hole in the core  of the
neutron star.  An  additional stage of accumulation  for bosons occurs
when      bosons       form      a       Bose-Einstein      condensate
\cite{McDermott:2011vn,goldman1989a,Kouvaris:2011fk}.   Finally,  { when dark matter
density $\rho_\chi$ is larger than the baryon density $\rho_b$,}
the  dark matter is self-gravitating,  namely, the
baryon density can be neglected. { Only during this  final stage of
collapse when self-gravity dominates is the bosonic  dark matter eventually  relativistic,  a requirement for
black hole formation.} 

\subsection{Dark matter accumulation regimes and characteristic times}

\subsubsection{Overview}

Dark  matter   accumulation  by   neutron  stars  occurs   in  several
stages. The first  stage is the capture of the  ambient dark matter by
the neutron star.  Cooling of  dark matter through interactions in the
neutron star cause the orbital radius  of the dark matter to decrease,
as dark matter continues to  accumulate. Thermalization of dark matter
with the neutrons  in neutron star, the possibility of  formation of a
Bose-Einstein condensate, the onset  of self-gravity and the potential
for the dark matter to coalesce into  a black hole are all elements of
the evolution  of the  accumulated dark matter.   In this  section, we
discuss  the capture  and  thermalization of  dark  matter by  neutron
stars.

Cooling   of   the   dark   matter  comes   from   interactions   with
neutrons.  Energy  is  transferred  to the  neutrons,  which  is  then
dissipated. To  first approximation,  we may treat  the neutrons  as a
kind of  static background for  dark matter accumulation  and cooling.
Scattering with  other dark  matter does  not promote  cooling because
there is little  dissipation of the dark matter.  Energy is transfered
between dark  matter particles,  contributing to dark  matter capture,
but not to dark matter evaporation from the neutron star. As discussed
by Zenter  in the Appendix of  Ref. \cite{Zentner:2009ys}, evaporation
of dark  matter particles  due to  dark matter-dark  matter scattering
will not occur  unless the escape velocity inside the  neutron star is
comparable to  the dark matter velocity  at a large distance  from the
neutron  star. A  typical value  for  the large  distance dark  matter
velocity is on the order of  $10^{-3}c$ in our galaxy at our location,
small compared  to the  escape velocity  of dark  matter in  a neutron
star.

\begin{table}[h] 
\centering
\begin{tabular}{cc}
  \hline
  \hline 
  Name & Value \\
  \hline
  Solar Mass & $M_{\odot} = 1.98892 \times 10 ^{33} $g \\
  Velocity dispersion of DM & $\bar{v}$ = 220 km/s \\
  Neutron Star Mass & M = 1.44 M$_{\odot}$ \\
  Neutron Star Radius & R = 10.6 km \\
  { Average} Density of Neutron Star & $\rho_{b} = 5.7\times10^{14} {\rm g/cm}^{3}$ \\
  Temperature inside the Neutron Star & 10$^{5}$ K\\
  \hline
\end{tabular}
\caption{Constants and parameters used in the results presented here.
}\label{constants}
\end{table}

\subsubsection{First phase - accumulation of dark matter by neutron star}

The initial accumulation  of dark matter by neutron  stars occurs with
dark matter  scattering with  nucleons with a cross section $\sigma_{\chi n}$ and  with energy  loss of  the dark
matter   of   approximately   
\be   \delta   E_\chi   \simeq   \frac{2
  m_r}{m_n+m_\chi} E_\chi \ee
 where  the average incoming kinetic energy is labeled $E_\chi=\langle T_{\rm  in}\rangle$ {and  the reduced  mass is
   $m_r=m_n  m_\chi/(m_n+m_\chi)$.}   In  the  constant  neutron  star
 density approximation, the average kinetic  energy of the dark matter
 inside  the neutron  star approximated  by a  trajectory through  the
 center of the neutron star is \cite{Kouvaris:2011uq} \be
\label{eq:t1}
\langle T_{\rm in}\rangle = \frac{4}{3} E_*+E_{\rm orbit}
\ee
where $E_*=GM m_\chi/R$ and $E_{\rm  orbit} = -G M m_\chi/(2 a)$
for  $a$  the  semimajor  axis.  {This  expression  follows  from  the
  assumption  that  the  density  of  the  neutron  star  $\rho_b$  is
  constant.}   Following Ref.  \cite{Kouvaris:2011uq}, for  $m_\chi\gg
m_n$ the  time $r_1$ for the orbit  to be contained within  the neutron star
radius   ($r_\chi(t_1)=R$)   is   
\be  t_1\sim   \frac{3\pi   m_\chi
  R^{3/2}\sigma_{crit}}{4   m_n\sqrt{2   G   M}   \sigma_{\chi   n}}\,
\sqrt{m_\chi/m_n} \ee 
where relativistic  corrections are neglected and  $\sigma_{crit}= m_n
R^2/M$.  Relativistic  corrections will not  change the time  scale by
more than a factor  of a few.  For a neutron  star with the parameters
of Table I, this leads to a ``containment time"
\begin{eqnarray}
\label{eq:tcontain}
t_1&\sim & 2.7\times 10^{-57}\Biggl(\frac{m_\chi}{m_n}\Biggr)^{3/2}\frac{\rm cm^2}{\sigma_{\chi n}}\ {\rm yr}\\
\nonumber &\simeq&
2.7\times 10^{-2}\Biggl(\frac{m_\chi}{m_n}\Biggr)^{3/2}\frac{1}{\sigma_{\chi n,55}}\ {\rm yr}\ ,
\end{eqnarray}
where  $\sigma_{\chi n,55}=  \sigma_{\chi n}/10^{-55}$  cm$^2$.  Other
choices  of  density distribution  of  neutrons  in  the NS  will  not
dramatically change $t_1$.  Note also that the expression for $t_1$ in
{ eq. (2.10) does not rely on the fact that the accumulation of the dark matter is specifically by a neutron star, only that  
$m_\chi>m_n$ and that using the average density of the star is reasonable in the evaluation of $t_1$ \cite{Kouvaris:2011uq}.}

Special to  neutron stars is  the effect  of the Pauli  suppression of
scattering. For  DM with  $m_\chi>m_n$ incident  on the  neutron star,
with $t<t_1$,  the characteristic  energy is  large enough  that Pauli
blocking does  not apply.   The Fermi energy  $E_F$ in the  zero temperature
approximation is
\begin{equation}
E_F=\frac{\hbar^2}{2m_n} (3\pi^2 n_b)^{2/3}\ ,
\end{equation}
where  $n_b$ is  the  number  density of  neutrons.   For the  average
neutron density used here, the Fermi energy is 97 MeV, which with $E_F
= p_F^2/2m_n$ determines the Fermi momentum $p_F\simeq 426$ MeV. Pauli
blocking  is represented  by the  factor  $\xi$ which depends on the change in
momentum $\delta p$,
\be \xi  = {\rm  min}\Biggl[
  \frac{\delta p}{p_F},1\Biggr]\ .  \ee 
For $m_r\simeq m_n<m_\chi$ and the  velocities relevant to this phase,
$\xi=1$. After $t_1$, Pauli blocking is important.

For $m_r <m_n$, { the evaluation of $t_1$ \cite{Kouvaris:2011uq} is modified by Pauli blocking} even for $t<t_1$, so that eq. (2.10) becomes
\be
t_1\simeq \frac{3\pi (m_n+m_\chi )R^{3/2}\sigma_{crit}}{4 m_r\sqrt{2 G M} \sigma_{\chi n}\xi }\, \sqrt{m_\chi/m_n}\ .
\ee
For this stage of the capture, we can take $\delta p \simeq \sqrt{2}m_r v_{esc}(R)$ where the escape velocity
at $r=R$ is $v_{esc}(R)=\sqrt{2 G M/R}
\simeq 0.63 \, c$.  For $m_\chi = 0.1$ GeV, this gives $t_1\sim 0.54\,  {\rm yr}/\sigma_{\chi n,55}$.

\subsubsection{Second phase - energy loss with orbits inside neutron star, neutron scattering dominated}

After $t\sim t_1$, the sphere of dark matter (the ``dark matter sphere") 
will have a radius of $r_\chi<R$. In this section we
discuss the evolution of $r_\chi$ with time. Again,  scattering of DM by neutrons gives a change in kinetic energy, $\delta E_\chi$, and as a result, the orbital radius decreases. We assume a circular orbit, as in Ref. \cite{Kouvaris:2011uq}. 
The kinetic energy can be expressed in terms of the  orbital radius $r_\chi(t)$, 
\be
\label{eq:er2}
E_\chi = \frac{2\pi }{3} G\rho_b m_\chi r_\chi^2,
\ee
with the time rate of change of the
kinetic energy,
\be
\label{eq:dedt}
\frac{dE_\chi}{dt} = -\xi n_b \sigma_{\chi n} v \delta E
\ee
with $\delta E = 2 m_r E_\chi/(m_n+m_\chi)$. Eqs. (\ref{eq:er2}) and
(\ref{eq:dedt}) can be combined to yield
\be
\frac{dr_\chi}{dt}=-\frac{4\pi\sqrt{2} }{3} \frac{m_r^3 n_b^2\sigma_{\chi n} Gr_\chi^3}{ m_\chi p_F}
\ee
for $t>t_1$. 
The solution is that the radius of the dark matter sphere is 
\begin{eqnarray}
\label{eq:rchit1}
r_\chi(t) &=& {R}{
\Biggl(1+\frac{8\pi\sqrt{2} m_r^3 n_b^2\sigma_{\chi n} G R^2 (t-t_1)}{3 m_\chi p_F}\Biggr)^{-1/2}}\\
\nonumber
 &&t>t_1,\ {\rm before\ thermalization}
\end{eqnarray}
for a constant neutron density in the neutron star. As already noted, the cooling which is responsible for the shrinking radius
comes from scattering of dark matter with neutrons. Dark matter scattering with accumulated dark matter will not contribute to
cooling, so eq. (\ref{eq:rchit1}) only depends on $\sigma_{\chi n}$, not $\sigma_{\chi\chi}$. Numerically, we find that
\begin{eqnarray}
r_\chi^2(t)&\simeq &\frac{m_\chi}{m_n} \frac{2.8\times 10^{10} \ {\rm cm^2 yr}}{\sigma_{\chi n,55} t}\quad m_\chi\gg m_n\ ,\\
r_\chi^2(t)&\simeq &\frac{m_n^2}{m_\chi^2} \frac{2.8\times 10^{10} \ {\rm cm^2 yr}}{\sigma_{\chi n,55} t}\quad m_\chi\ll m_n\ .\\
\end{eqnarray}

The next phase in dark matter accumulation may include a significant time in which $\chi-\chi$ scattering dominates $\chi-n$ scattering.
Since only $\chi-n$ scattering is relevant to cooling, we can already
evaluate the thermalization time.
The thermal radius $r_{th}=r_\chi(t_{th})$ is related to the neutron star temperature $T$ in the core
via
\be
E_\chi = \frac{2\pi}{3} G\rho_b m_\chi r_{\chi}^2 (t_{th})=\frac{3}{2}kT\ .
\ee
With the solution for $r_\chi(t)$ in eq. (\ref{eq:rchit1}) at $t=t_{th}$, for $t_{th}\gg t_1$, one recovers the usual expression
(see, e.g., eq. (19) in Ref. \cite{McDermott:2011vn}), 
\begin{eqnarray}
t_{th} &= &\frac{m_\chi^2 p_F}{6\sqrt{2}T m_n^2 n_b\sigma_{xn}}\cdot\frac{m_n^3}{m_r^3}\\ \nonumber
&\simeq& 2.5\times 10^5 \ {\rm yr} \Biggl(\frac{m_\chi}{m_n}\Biggr)^2 \Biggl(\frac{m_n}{m_r}\Biggr)^3
\frac{1}{\sigma_{\chi n,55}}\ ,
\end{eqnarray}
for $T=10^5$ K,  a constant neutron star density and $m_\chi>m_n$.
Numerically, the thermal radius is 
\be
r_{th} = \Biggl(\frac{9kT}{4\pi G\rho_b m_\chi}\Biggr)^{\frac{1}{2}}=334\ {\rm cm}\Biggl(\frac{m_n}{m_\chi}\Biggr)^{\frac{1}{2}} 
\ee

\subsubsection{The onset of self-capture}

For some sets of parameters, the third phase of dark matter capture by the neutron star comes from the effects of dark matter
interactions with the dark matter already captured by the neutron star. 
In the next section, we discuss the form of the neutron capture and self-interactions of dark matter for { the number of accumulated
dark matter particles} $N_\chi(t)$.
The neutron capture  { of dark matter} depends on  { the cross section and number of target neutrons
$N_n$ via } $\sigma_{\chi n} N_n$ while the self capture term for $dN_\chi/dt$ depends
on { the self-interaction cross section and the number of already accumulated dark matter particles $N_\chi$ via} 
$\sigma_{\chi\chi} N_\chi$. When self-capture dominates, the linear growth of $N_\chi(t)$ changes to an exponential growth.
We label  the
time at which the dark matter begins to accumulate
exponentially by $t_2$.

When the target dark matter is
effectively
at rest, the change in DM kinetic energy from $\chi \chi$ scattering is
\be
0\le \frac{\delta E_\chi}{E_\chi}\le 1 \ .
\ee
The kinetic energy of a dark matter particle within the neutron star at
orbital radius $r_\chi$ is
\be
E_\chi = \frac{2\pi}{3} G\rho_b m_\chi r_\chi^2
=\frac{1}{4} m_\chi \Biggl(v_{esc} (R) \frac{r_\chi}{R}\Biggr)^2\ ,
\ee
and  $\delta   E_\chi\sim  0.5  E_\chi$  for   $\chi\chi$  scattering,
neglecting  the  recoil  of  the   DM  that  resides  in  the  neutron
star. Recoil is important when evaporation of dark matter particles is
a possibility. As discussed in Ref. \cite{Zentner:2009ys}, the recoils
can be  neglected when the  dark matter escape speeds  are significant
compared to the  average speed of the (neutron) star  and the velocity
dispersion in the  dark matter halo.  The escape speed  on the surface
of a neutron  star is of order 0.6$c$, and  the escape speed increases
as  the dark  matter goes  further into  the interior  of the  neutron
star. By comparison, the solar speed in  the galaxy is on the order of
220 km/s and the DM velocity dispersion is 270 km/s, several orders of
magnitude smaller  than the dark  matter escape speed.   Therefore, we
can use eq.  (\ref{eq:rchit1}) for the radius of the  wimp-sphere as a
function of time, even when $\chi \chi$ scattering is important.

An exponential  increase in the number  of dark matter particles  as a
function of time because of  self-interactions makes the formation of a
black hole from dark matter a  possibility, even in the regime of slow
thermalization  considered  here.  There is  an  important  mitigating
effect,  however, coming  from the  geometric limit.   The exponential
growth of $N_\chi(t)$ is cut off  when the geometric limit is reached, 
\be
N_\chi (t_G)\sigma_{\chi\chi} = \pi r_\chi^2(t_G)\ , 
\ee
{ where $t_G$ is the time at which eq. (2.27) is satisfied.  }
In  practice, the  thermalization time  occurs before  $t_G$ for  some
parameter choices, in which case  eq. (\ref{eq:rchit1}) does not apply
after $t>t_{th}$.

\subsection{Conditions for relativistic dark matter particles}

With thermalization, for fermionic dark matter, one can use $r_{th}$ in
eq. (\ref{eq:nfrel}) to find that
\begin{equation}
N_f\geq N_f^{rel}(r=r_{th})=1.9\times 10^{51}\Biggl( \frac{m_\chi}{100\ {\rm GeV}}
\cdot\frac{T}{10^5K}\Biggr)^{3/2}\ ,
\end{equation}
in order for fermionic dark matter to be relativistic.  For $T=10^5K$,
the  Chandrasekhar limit  for  fermions,  which requires  relativistic
particles, is  applicable only  for dark matter  masses below  100 GeV
when      $t>t_{th}$,      since       for      $m_\chi<100$      GeV,
$N_f^{Ch}>N_f^{rel}$.  When $m_\chi>100$  GeV,  fermionic dark  matter
within     the    thermal     radius     with    $N_f=N_f^{Ch}$     is
non-relativistic. This appears not to  have been taken into account in
Ref. \cite{deLavallaz:2010wp}.

\begin{figure}  
\centering
   \includegraphics[angle=270,scale=0.4]{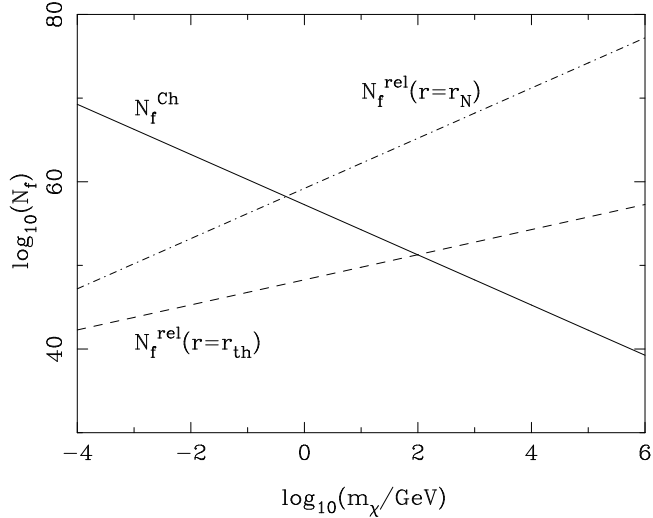}
   \caption{As a function of fermionic dark matter $m_\chi$, log$_{10}$ of the number
   of fermions for the Chandrasekhar limit (which requires the dark matter to be relativistic)
   shown with the solid line and
   the minimum number of fermions  required for relativistic energies when
   $r=R$ (dot-dashed) and $r=r_{th}$ (dashed).}
\label{fig:nf}
\end{figure}

Fig.  \ref{fig:nf}  shows the  relevant  $N_f$  for the  Chandrasekhar
limit, for  $N_f^{rel}(r=r_{th})$ and the number  of fermions required
for  relativistic fermions  when $r=R$,  the radius  of the  neutron
star.  Before  the thermalization time,  the larger of  $N_f^{Ch}$ and
$N_f^{rel}(r=R)$ determines whether or  not the dark matter fermions
collapse  to a  black  hole.  For $t>t_{th}$,  it  is  the maximum  of
$N_f^{Ch}$ and $N_f^{rel}(r=r_{th})$.

If the dark  matter particles are bosons, there  are additional stages
of accumulation  of dark  matter particles. Within  the thermalization
radius, as  the number of  bosons continues  to increase in  time, the
bosons   can   further   form   a   Bose-Einstein   condensate   (BEC)
\cite{goldman1989a}. The  particle number  density required to  form a
BEC  in  a   sphere  of  radius  $r_{th}$  in  the   neutron  star  is
\cite{goldman1989a,Kouvaris:2011fk}
\begin{equation}
N_{b}^{BEC}(r=r_{th}) =  2.7\times 10^{36} \Biggl( \frac{T}{10^5\ K}\Biggr)^3\ .
\label{bec} 
\end{equation}
Once the BEC  forms, the radius in which the  dark matter particle can
reside, which is  the radius of the wave function  of the ground state
in  the  gravitational potential  of  the  neutron star  becomes  even
smaller                          than                         $r_{th}$
\cite{goldman1989a,Kouvaris:2011fk,McDermott:2011vn}, namely,
\begin{eqnarray}
\nonumber
r_{BEC} &=& \left(\frac{3}{8 \pi G m^2_{\chi} \rho_{b}}\right)^{1/4} \\
&=& 1.5\times 10^{-5}\ {\rm cm}\Biggl(\frac{100\ {\rm GeV}}{m_\chi} \Biggr)^{1/2}\  .
\label{rbec}
\end{eqnarray}
Clearly, $r_{BEC} < r_{th}$, which has significant implications 
for reaching the Chadrasekhar limit in less time.  


If  the  dark  matter  particles  are bosons,  they  can  only  become
relativistic  and  collapse  if they  become  self-gravitating.   This
happens  when the  density of  the dark  matter particles  exceeds the
baryon density within the same  volume. Since the nucleon density does
not change in time in the neutron  star, as soon as the number of dark
matter  particles  reaches  a   critical  number,  the  bosons  become
self-gravitating. This critical number is given by
\begin{equation}
\label{eq:nselfgeneral}
N_{self} = \left(\frac{4 \pi r^{3} \rho_{b}}{3 m_{\chi}}\right) \ .
\end{equation}
For $r=r_{th}$,
\begin{equation}
N_{self}(r=r_{th})\simeq 4.8 \times 10^{46} \left(\frac{1~{\rm
    GeV}}{m_{\chi}}\right)^{5/2} \left(\frac{T}{10^5 ~{\rm K}}\right)^{3/2}
\  .
\label{eq:ngrav}
\end{equation}
Clearly,  as  the  radius  of  the dark  matter  core  decreases,  the
condition  for the  dark matter  particles to  become self-gravitating
will be easier to reach, which  would eventually cause the dark matter
core  to  collapse and  form  a  black  hole. When  the  Bose-Einstein
condensate has formed,
\begin{eqnarray} 
N_{self}(r=r_{BEC})\simeq 1.1 \times 10^{28} \left(\frac{1~{\rm
    GeV}}{m_{\chi}}\right)^{5/2}
\ .
\label{ngravbec}
\end{eqnarray}
    
\begin{figure}
\centering
   \includegraphics[angle=270,scale=0.4]{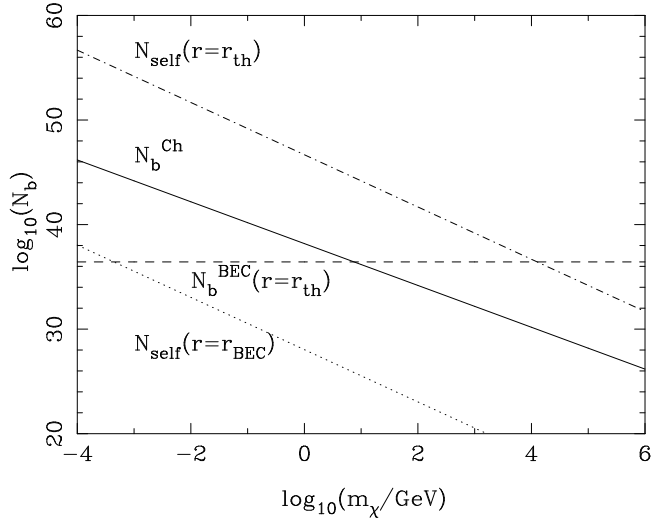}
   \caption{As a function of bosonic dark matter $m_\chi$, log$_{10}$ of the number
   of bosons for the Chandrasekhar limit (which requires the dark matter to be relativistic)
   shown with the solid line,
   the minimum number of bosons  required for the bosons to be self-gravitating when
   $r=r_{th}$ (dot-dashed) and $r=r_{BEC}$ (dotted), and the number of bosons
   within the radius $r=r_{th}$ for a Bose-Einstein condensate to form (dashed).}
\label{fig:nb}
\end{figure}

Comparing the equations for  $N_b^{Ch} $ (eq. (\ref{chandra_bos})) and
for  self-gravitation in  eq. (\ref{eq:ngrav})  one can  see that  the
Chandrasekhar limit  is already exceeded for  thermalized bosonic dark
matter    when   self-gravity    is   established,    as   shown    in
Fig.  \ref{fig:nb}.   Self-gravity  results  in  further  in-fall  and
ultimately   relativistic   bosons,    the   final   requirement   for
Eq. (\ref{chandra_bos}) to apply.

To summarize, the necessary condition  for collapse of neutron star to
a black hole is that the dark  matter is thermalized within the age of
the neutron star, which we take  to be $10^9$ years.  Additionally, in
the absence of  a specific calculation of the  Chandrasekhar limit for
bosons  with  strong  self-interactions,  we  require  the  number  of
captured  dark  matter  particles  to  exceed  the  following  limits:
$N_b^{Ch}$ for $m_\chi < 7.4$ GeV,  $N_b^{BEC}$ for $ 7.4 \ {\rm{GeV}}
< m_\chi < 1.26 \times 10^4$ GeV, and $N_{self}(r=r_{th})$ for $m_\chi
> 1.26 \times 10^4$ GeV.

\section{Evaluation of Capture of Dark Matter Particles by Neutron Stars}


The time evolution of the dark matter particles
captured by a neutron star is given by
\cite{Zentner:2009ys}
\begin{equation}
\frac{dN_{\chi}}{dt} = C_{c}+C_{s}N_{\chi}-C_{a}N_{\chi}^{2},
\label{maineq}
\end{equation}
where  $C_{c}$  is  the  capture   rate  due  to  dark  matter-nucleon
interactions, $C_{s}N_\chi$ is the capture rate due to the dark matter
self-interactions  and  $C_{a}N^{2}_{\chi}$   governs  the  number  of
particles lost due to the their annihilation.  We consider the case of
asymmetric dark matter, when dark  matter particles do not annihilate,
{ so $C_a=0$.  In this case, for time-independent $C_c$ and $C_s$, the
  solution for the eq. (\ref{maineq})  is }
\beq N^0_\chi = \frac{C_c}{C_s}(e^{C_s
    t} -1) \ .
\label{simplen}
\eeq
{ Here the notation $N_\chi^0$ (rather than $N_\chi$) is 
to indicate that this is the solution
when $C_c$ and $C_s$ are time independent.}
{We consider times up to $t_{max}=10^9$ yr.}

The  dark matter  self interaction  cross section  $\sigma_{\chi\chi}$
enters linearly in $C_s$, as discussed  below in Sec. 3.2.  {As noted
  above,  when $N_\chi\sigma_{\chi\chi}>\pi  r_\chi^2$, the  geometric
  cross section  for $\chi\chi$ scattering  in the neutron  star, both
  eq.  (\ref{maineq})  and  eq. (\ref{simplen})  are  modified.  These
  modifications are outlined  in Sec. 3.2, and we show  an example of
  how the  geometric limit on $\chi$  interactions with the DM  in the
  neutron star affects the growth of $N_\chi$ as a function of time.}

\subsection{Dark matter-nucleon interactions}
\label{dmn_sec}

The number of dark matter particles  that can be captured by a neutron
star is given by \citep{Kouvaris:2008zr} 
\begin{eqnarray}
\nonumber
C_c&=&\frac{8}{3} \pi^2 \frac{\rho_{\chi}}{m_{\chi}} \left ( \frac{3}{2 \pi
 \bar{v}^2} \right )^{3/2} \frac{GMR}{1-\frac{2GM}{R}}
\bar{v}^2(1-e^{-3\epsilon_0/\bar{v}^2})\\
&\times & \xi\,\,  f ~\rm{particles ~ s^{-1}},
\label{accretion}
\end{eqnarray}
where  the  general  relativistic  affects  assuming  a  Schwarzschild
geometry have been incorporated. Here $\rho_{\chi}$ is the dark matter
density at the location of the neutron star, $m_{\chi}$ is the mass of
the dark matter particles, $M$ and $R$  are the mass and the radius of
the neutron star, respectively, and  $\bar{v}$ is the average velocity
of dark matter particles  in the Galactic halo \cite{Kouvaris:2010kx}
{ and $\xi$ is the Pauli blocking factor defined in eq. (2.13).}
For a  dark matter  particle to  be trapped  with a  single collision,
$E<m_\chi    \epsilon_0$,    which   defines    $\epsilon_0$.    Since
$\epsilon_0\gg \bar{v}^2/2$ we take $e^{-3 \epsilon_0/\bar{v}^2}\simeq
0$ \cite{Kouvaris:2008zr}.   In order  for a  dark matter  particle to
stay inside a  neutron star it must lose enough  energy by interacting
with the  particles inside the neutron  star.  The last factor  in eq.
(\ref{accretion}),  $f$, determines  the fraction  of the  dark matter
particles  that  will  be  trapped  inside the  neutron  star  due  to
interactions   with   nucleons,   which   for   $\sigma_{\chi   n}   <
\sigma_{\text{crit}}=     \pi     m_nR^2/M     \simeq     2     \times
10^{-45}~\text{cm}^2$, $f$ can be approximated as \beq f=\Big{\langle}
1-                                  \exp\Big{[}-\int\frac{\sigma_{\chi
      n}\rho}{m_n}dl\Big{]}\Big{\rangle}\simeq
\Big{\langle}\int\frac{\sigma_{\chi n}\rho}{m_n}dl \Big{\rangle}\ ,
\label{f1}
\eeq
where $dl $ is the infinitesimal arc length of a trajectory in the
neutron star \cite{press1985a,Kouvaris:2008zr}. 
Assuming constant density for the neutron star, numerical evaluation 
of $f$ in terms of the 
critical cross section, 
$\sigma_{\rm crit}$, which is the geometrical cross section of the neutron 
star divided by the
number of nucleon targets, is given by 
 \cite{Kouvaris:2008zr}
\beq
   f                             \simeq 
\frac{\sigma_{\chi n}}{\sigma_{\text{crit}}}\Big{\langle}\int\frac{\rho}{M/R^3}\frac{dl}{R}\Big{\rangle}
\simeq 0.45 \frac{\sigma_{\chi n}}{\sigma_{crit}}\ .
\label{f1}
\eeq
This equation is applicable to the case where
$\sigma_{\chi  n} <  \sigma_{\rm crit}$,  however $f$  saturates to unity
when $\sigma_{\chi n}$ is larger than $\sigma_{\rm crit}$
\cite{Kouvaris:2008zr}.  Thus, the quantity $f$ is
\beq
f \simeq  0.45 \frac{{\rm min}(\sigma_{\chi n},\sigma_{crit})}{\sigma_{crit}}\ .
\label{f}
\eeq

Currently, the  most stringent  experimental limits on  the DM-nucleon
cross section, $\sigma_{\chi n}$, come from the XENON \cite{xenon} and
CDMS  \cite{CDMS} experiments  which  look for  energy deposition  via
nuclear recoils  from dark  matter scattering.   For dark  matter mass
between 20 GeV and $100$ GeV, the limit is $\sigma_{\chi n} < 6 \times
10^{-44}$ cm$^2$, while for larger  mass, i.e. $m_\chi= 10^3$ GeV, the
limit is  about an order of  magnitude larger and similar  for lighter
dark matter particle, when $m_\chi  \sim 15$ GeV.  Recently, CRESST-II
has claimed $4.7\sigma$ signal corresponding  to $m_\chi \sim 10 - 30$
GeV  and  for  $10^{-40}\  {\rm  cm}^2 <  \sigma_{\chi  n}<  3  \times
10^{-43}$cm$^2$  \cite{CRESST}.   This  signal  does not  seem  to  be
consistent  with  DAMA  results   claiming  annual  modulation  effect
consistent with dark matter particle of  mass around $10$ GeV with the
cross section  $\sigma_{\chi n} \sim 10^{-40}$cm$^2$  \cite{DAMA}.  We
will  incorporate these  limits  by taking  conservative approach  and
considering  only cases  when  $\sigma_{\chi n}$  is below  $10^{-44}$
cm$^2$, {written as  $\sigma_{\chi n}=\sigma_{\chi n,55}10^{-55}\ {\rm
    cm}^2$.}

The parameters that  we use are given in  Table \ref{constants}.  With
these  parameters, the  capture rate,  $C_c$, for  $\sigma_{\chi n}  <
\sigma_{crit} = 2 \times 10^{-45}$ cm$^2$, is given by,
\beq
C_c = 9.19 \times 10^{22} \frac{\sigma_{\chi n,55}}{m_\chi /{\rm GeV}} 
\frac{\rho_\chi}{{\rm GeV/cm^3}} \xi \,\, 
{\rm yr^{-1}} 
\eeq

\subsection{Dark Matter Self Interactions}
\label{dmm_sec}

Once  a certain  amount of  dark matter  particles accumulates  in the
neutron star, their very existence inside the neutron star will affect
the capture of new dark matter particles due to their self-interaction
\cite{Zentner:2009ys}.  The  self-capture rate,  C$_{s}$, is  given by
\cite{Zentner:2009ys}
\begin{equation}
C_s      =     \sqrt{\frac{3}{2}}      \frac{\rho_{\chi}}{m_{\chi}}     \sigma_{\chi \chi}
v_{esc}(R)\        \frac{v_{esc}(R)}{\bar{v}}       \langle
\hat{\phi}_{\chi} \rangle \frac{\erf(\eta)}{\eta}\frac{1}{1-\frac{2GM}{R}},
\label{C_s}
\end{equation}
where,  $\sigma_{\chi \chi}$  is  the dark  matter elastic  scattering
cross-section and $v_{esc}(R)$ is the escape velocity from the surface
of the neutron  star.  As discussed in Sec. 2.2,  it is primarily the
gravitational effect  of the neutron  star that keeps the  dark matter
inside the neutron star, eventually  thermalizing, so eq. (\ref{C_s}) depends
on the  neutron star radius  $R$ rather than  the details of  the dark
matter distribution within the neutron  star.   We  modified eq.  (\ref{C_s})
following Ref.  \cite{Kouvaris:2008zr} in  order to take  into account
the general  relativistic effects assuming as  Schwarzschild geometry.
The quantity $\hat{\phi}_{\chi}$ which signifies how compact the star,
is a  dimensionless potential  defined as  \cite{Zentner:2009ys},
 \beq
\hat{\phi}_{\chi} = \frac{v_{esc}^{2}(r)}{v^{2}_{esc}(R)}\  .  \eeq
 We
take  $\langle\hat{\phi}_\chi\rangle  =  1$,  which is  valid  in  the
approximation that  the mass density  of the neutron star  is uniform.
Since  the  density  is  larger   in  the  core,  which  implies  that
$\langle\hat{\phi}_\chi\rangle >  1$, our assumption  is conservative.
Finally, $\eta^2\equiv 3/2(v_{N}/\bar{v})^2 $  depends on the velocity
of  the  neutron star  $v_N$  in  the  Galaxy.  We  approximate  ${\rm
  erf}(\eta)/\eta\simeq 1$.

The most stringent limits on self-interaction cross section, 
$\sigma_{\chi \chi}$, come 
from the 
Bullet  Cluster observations \cite{markevitch2004a}, i.e.
\begin{equation}
 \frac{\sigma_{\chi\chi}/10^{-24}\ {\rm cm}^2}{m_{\chi}/{\rm GeV}}<2\ ,
\label{dm_inter_limit}
\end{equation}
{so we scale the self interaction cross section, $\sigma_{\chi \chi} = \sigma_{\chi\chi, 24}10^{-24}\ {\rm cm}^2$.}
With our choice of 
the parameters given in Table I, 
$C_s$ is given by, 
\beq
C_s = 1.06 \times 10^{-3} \frac{\sigma_{\chi \chi, 24}}{m_\chi / {\rm GeV}}\frac{\rho_\chi}{{\rm GeV/cm^3}}
{\rm yr^{-1}}\ .
\eeq

\begin{figure}
\centering 
   \includegraphics[angle=270,scale=0.4]{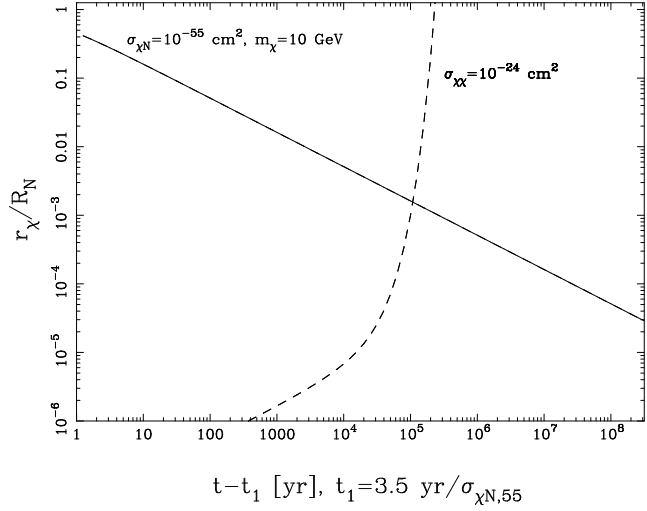}
   \caption{As a function of time, $r_\chi(t)/R$ (solid line) and $\sqrt{N^0_\chi(t)\sigma_{\chi\chi}/\pi R^2}$ 
(dashed curve) for $\sigma_{\chi\chi,24}=1$ and
$\sigma_{\chi n,55}=1$, $m_\chi=10$ GeV and $\rho_\chi=1$ GeV/cm$^3$. The time at which the solid line and dashed curve
cross is $t_G$.} 
\label{fig:tgexample}
\end{figure}

\begin{figure}
\centering 
   \includegraphics[angle=270,scale=0.4]{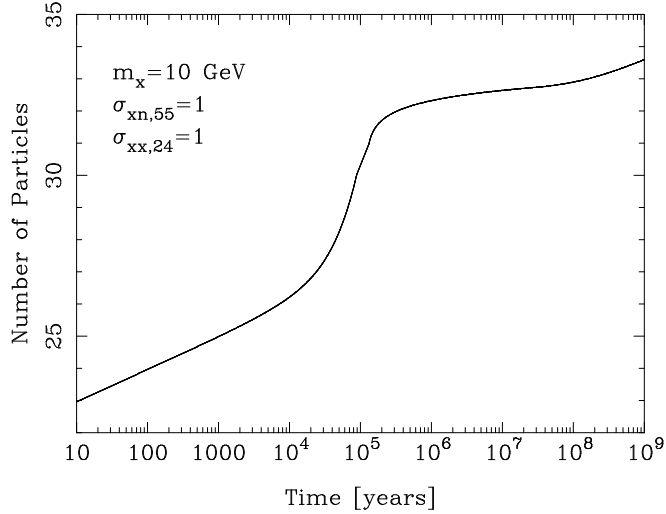}
   \caption{As a function of time, $N_\chi(t)$ for $\sigma_{\chi\chi,24}=1$,
$\sigma_{\chi n,55}=1$, $m_\chi=10$ GeV and $\rho_\chi=1$ GeV/cm$^3$.} 
\label{fig:nx1011}
\end{figure}

{The full  evolution of  $N_\chi(t)$ depends on  how quickly  the dark
  matter  thermalizes,  and whether  or  not  the geometric  limit  is
  reached,  before  or after  thermalization.   We  first discuss  the
  circumstance where $t_{G}<t_{th}<t_{max}$.   As discussed above, the
  geometric  limit for  DM capture  via $\chi\chi$  interactions, when
  $N_\chi(t_G)<\pi    r_\chi^2(t_G)/\sigma_{\chi\chi}$,   halts    the
  exponential   increase    in   DM   accumulation.     For   $t<t_G$,
  eq.   (\ref{maineq})  with   $C_a=0$,  \be   \frac{dN_{\chi}}{dt}  =
  C_{c}+C_{s}N_{\chi},
\label{maineqa}
\ee
governs the evolution. Between
$t_{G}<t<t_{th}$, the equation governing the
time evolution of $N_\chi$ is
\begin{eqnarray}
\label{dmspheredt}
\frac{dN_{\chi}}{dt} &=& C_{c}+C_{s}N_{\chi}(t_G)\times \frac{r_\chi^2(t)}{r_\chi^2(t_G)}\\ \nonumber
&\simeq&C_{c}+C_{s}N_{\chi}(t_G)\times \frac{t_G}{t} \ .
\end{eqnarray}
The shrinking dark matter sphere accounts for the second term in eq. (\ref{dmspheredt}).
Once the thermalization time is reached, a third phase of dark matter accumulation occurs, with
\be
\frac{dN_\chi}{dt} &=& (C_c+C_s \frac{\pi r_{th}^2}{\sigma_{\chi\chi}}),\quad t_{th}<t<t_{max}\ .
\ee

The solutions to these equations give the number of dark matter particles at $t=t_{max}$ of
\begin{eqnarray}
\nonumber
N_\chi (t_{max} )&=&N_\chi(t_{th})\\ 
\label{eq:Ntmax}
&+&\Bigl(C_c+C_s\frac{\pi r_{th}^2}{\sigma_{\chi\chi}}  \Bigr) (t_{max}-t_{th})\\ \nonumber
N_\chi(t_{th}) &=& N_\chi(t_G) + C_c(t_{th}-t_G) \\
\label{eq:Nth}
 &+&C_s N_\chi(t_G) t_G \ln \Biggl(\frac{t_{th}}{t_G}\Biggr)\\
\label{eq:Ntg}
N_\chi(t_G) &=& \frac{\pi r_\chi^2 (t_G)}{\sigma_{\chi\chi}}\ .
\end{eqnarray}

As an illustration, we show in Fig. \ref{fig:tgexample} the evolution of $r_\chi(t)/R$ 
and $\sqrt{N^0_\chi(t)\sigma_{\chi\chi}/\pi R^2}$ for $\sigma_{\chi\chi,24}=1$  with
$\sigma_{\chi n,55}=1$ and $m_\chi=10$ GeV. { The quantity $N_\chi^0(t)$ is defined
in eq. (3.2).} 
We also use $\rho_\chi = 1$ GeV/cm$^3$, equivalent to a distance of $d=3.7$ kpc from the Galactic
Center for a dark matter density parameterized by the Navarro-Frank-White halo profile \cite{Navarro}.
The intersection of lines is $t_G$. For $\sigma_{\chi\chi,24}=1$,
$t_G\simeq 10^5$ yr.
The resulting time evolution of $N_\chi$ for $\sigma_{\chi\chi,24}=1$ is shown in Fig. \ref{fig:nx1011}.}

{
For other parameters, the time ordering of $t_G$, $t_{th}$ and $t_{max}$ can be different.
If $t_{th}<t_G,\ t_{max}$, then the dark matter sphere decreases to the thermal radius while
the number of accumulated dark matter particles follows $N_\chi^0(t)$, eq. (3.2). After $t_{th}$, 
at some point $t_G'$,
\be
N_\chi(t_G')\sigma_{\chi\chi} = \pi r_{th}^2\ .
\ee
when the geometric limit is reached.
If $t_G'>t_{max}$ then
\be
\label{eq:Ntmaxpp}
N_\chi(t_{max})=\frac{C_c}{C_s}\Biggl(e^{C_s t_{max}} -1\Biggr)\ .
\ee
If $t_{th}<t_G'<t_{max}$ then
\be
\label{eq:Ntmaxp}
N_\chi (t_{max} )= \frac{\pi r_{th}^2}{\sigma_{\chi\chi} } +\Biggl(
\frac{C_s\pi r_{th}^2}{\sigma_{\chi\chi}}+C_c\Biggr)(t_{max}-t_G')\ .
\ee
}

\section{Results and Discussion}

In Figures  \ref{f1gev}, \ref{f10gev} and \ref{f100gev},  we show the
time  evolution of  the number  of  dark matter  particles  captured by  a
neutron star for 
$m_\chi = 1$ GeV, 
$m_\chi = 10$ GeV and 
$m_\chi = 100$ GeV, and for 
several values of dark matter interaction cross sections.  
We have taken the
Galactic  dark matter density to be 1 GeV~cm$^{-3}$
and  other parameters as given  in Table  \ref{constants}. 
The self-interaction cross section is taken to be 
$\sigma_{\chi \chi}=10^{-24}$ cm$^2$ and we consider two different 
values for dark matter nucleon interaction cross section, 
$\sigma_{\chi n}=10^{-55}$ cm$^2$ 
(top panels of Figs. 4-6) and 
$\sigma_{\chi n}=10^{-48}$ cm$^2$ 
(bottom panels of Figs. \ref{f1gev}-\ref{f100gev}).  
When the dark matter-nucleon cross section is small, $\sigma_{\chi n}=10^{-55}$ cm$^2$,
the dark matter self-capture is important, as shown by the upper solid curve in the top panels of Figs. \ref{f1gev}-\ref{f100gev}. 
Neglecting self-capture 
yields the lower dashed lines in the top panels.  In the bottom panels, 
$\sigma_{\chi n}=10^{-48}$ cm$^2$ and the self-capture contribution is negligible
compared to the capture by nucleons.

Figs. \ref{f1gev}-\ref{f100gev}  illustrate features of the evolution of $N_\chi(t)$ for low $\sigma_{\chi n}$ and large
$\sigma_{\chi\chi}$. When self-interactions are important, the accumulated dark matter in neutron stars can be several
orders of magnitude larger than without self-interactions. Is $N_\chi(t_{max})$ sufficiently large, with self-interactions, to
potentially form a black hole and disrupt the neutron star?
For $m_\chi=10$ GeV, our minimal limit is $N_b^{BEC}(r=r_{th})=2.7\times 10^{36}$ for
$T=10^5$ K, a limit larger by almost three orders of magnitude than $N_\chi(t_{max})$ when $\rho_\chi=1$ GeV/cm$^3$.
This leads us to consider regions where $\rho_\chi$ is on the order of $10^3$ GeV/cm$^3$, since
to a good approximation, $N_\chi(t)$ scales linearly with $\rho_\chi$. 

{ 
Assuming that the dark matter density in the Galaxy follows the NFW distribution, it can be seen that regions where the DM density can reach to $\rho_{\chi} \sim 10^3$ GeV/cm$^3$ are very limited. However, globular clusters may provide the regions with excessive DM density. 
 It has been shown by Bertone and Fairbairn \cite{Bertone:2007ae} 
 that within the core radius of the globular cluster M4, the 
 DM density may reach to $\approx 34,000$ GeV/cm$^3$, 
 although the existence of large dark matter densities 
 associated with globular
clusters has been controversial \cite{Conroy:2010bs}.  
Within the core of this cluster, 
the well known millisecond pulsar PSR~B1620-26 has been detected \cite{lyne}. 
This pulsar has a minimum 
characteristic age of $2.2\times 10^{8}$ yr \cite{mckenna}.  

Similar to M4, many neutron stars have been discovered in other globular clusters. For example, the globular cluster 47 Tuc has a core radius of 0.6 pc. In the core of the cluster at least 5 neutron stars have been located. In total, within a few times the core radius of this cluster, 20 neutrons stars have been located all with typical minimum characteristic ages of a few $\times10^{8}$ yr \cite{freire}.  
So while $\rho_\chi=10^3$ GeV/cm$^3$ is likely not common, 
it appears that there might be regions with such high
dark matter densities with old neutron stars.

\begin{figure*}
\centering
   \includegraphics[angle=270,scale=0.38]{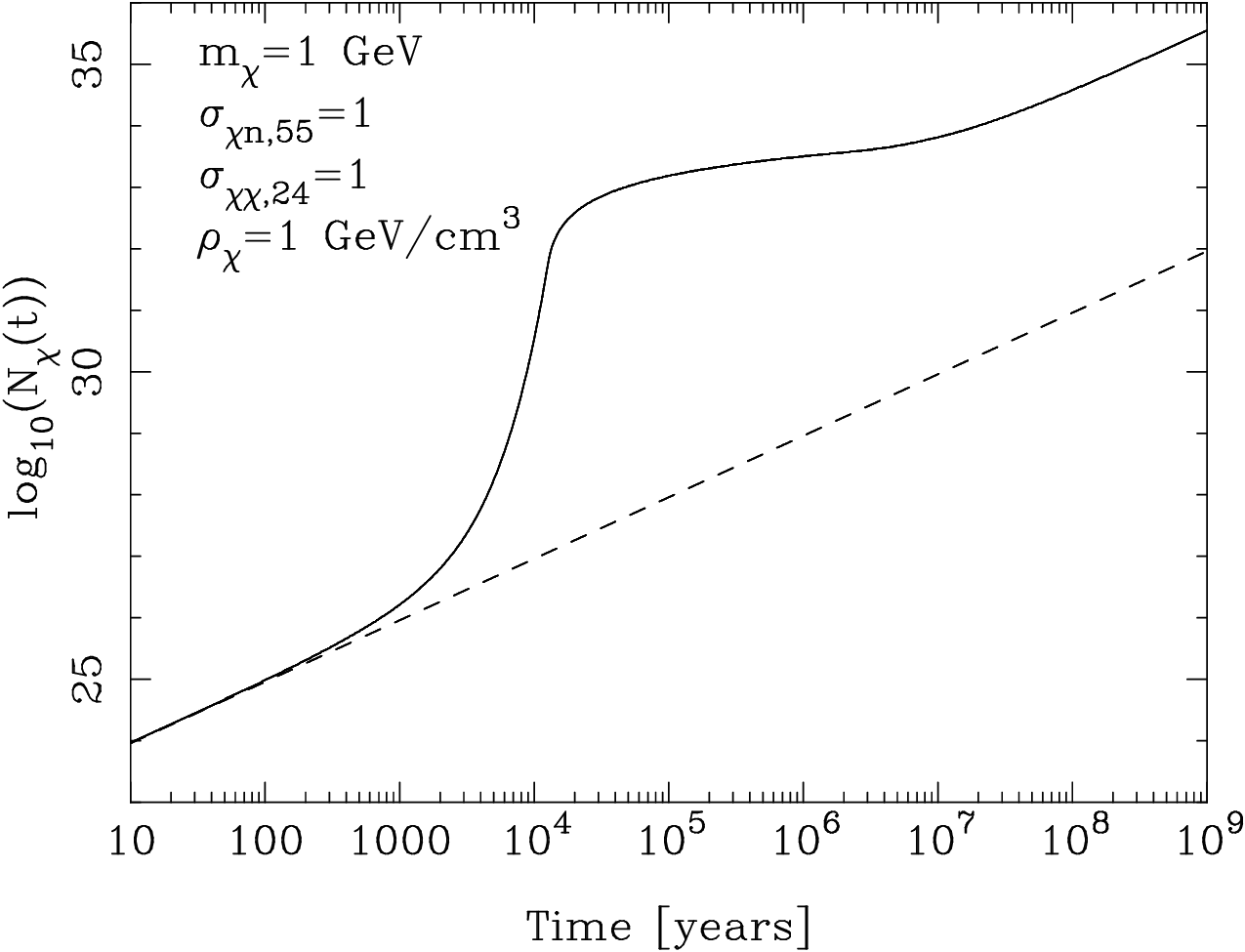}
   \includegraphics[angle=270,scale=0.38]{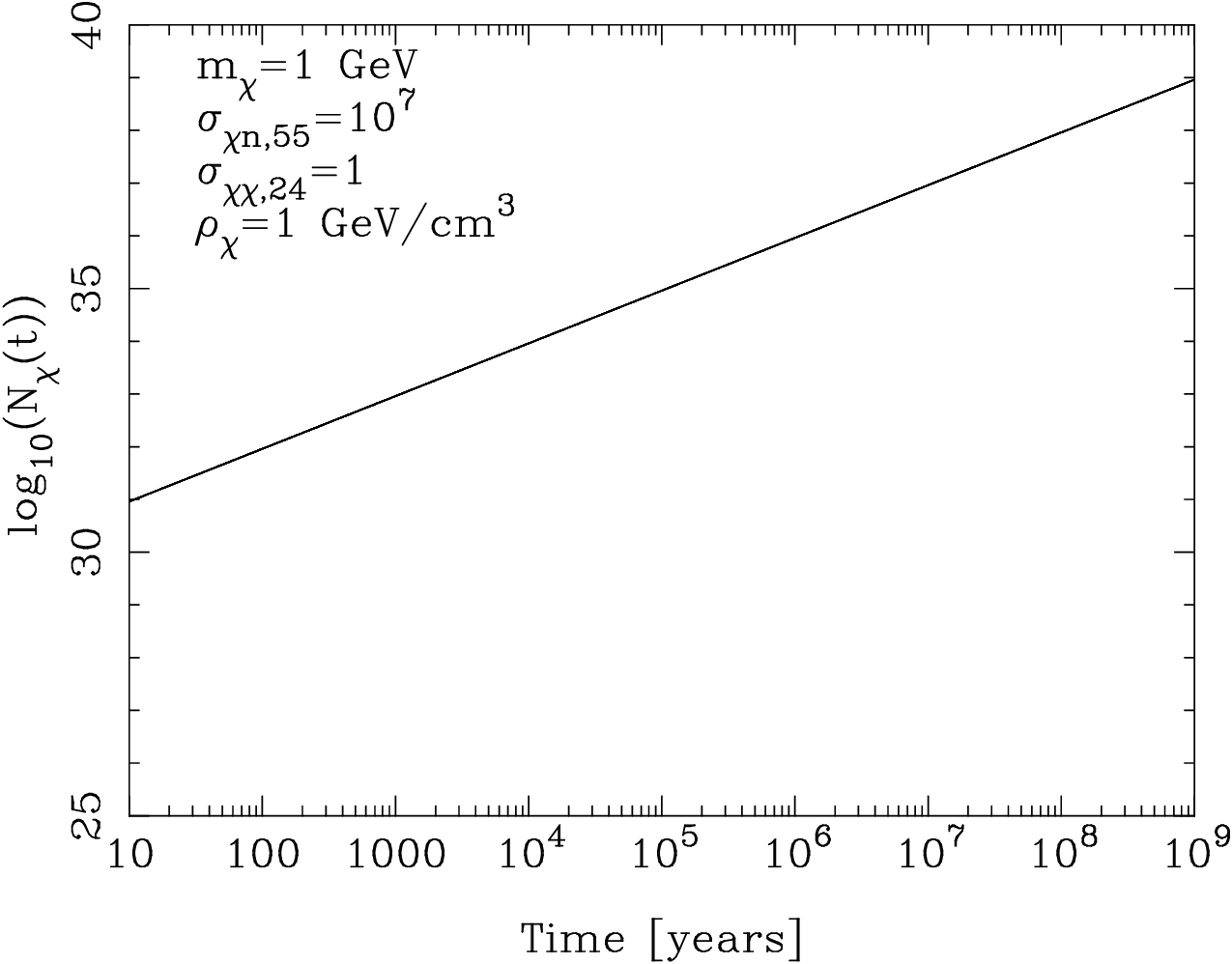}
   \caption{Time evolution of the number of asymmetric bosonic dark 
 matter particles accumulated in a neutron star, for 
     $m_\chi=1$ GeV,  
     $\sigma_{\chi \chi} = 10^{-24}$ and 
  $\sigma_{\chi n} = 10^{-55}$ cm$^{2}$ (top panel)  
    and 
  $\sigma_{\chi n} = 10^{-48}$ cm$^{2}$ (bottom panel).  The DM density is taken to be
$\rho_\chi=1$ GeV/cm$^3$. The dashed line in the top panel is for $\sigma_{\chi\chi,24}=0$.
} 
\label{f1gev}
\end{figure*}

\begin{figure*}
\centering
   \includegraphics[angle=270,scale=0.38]{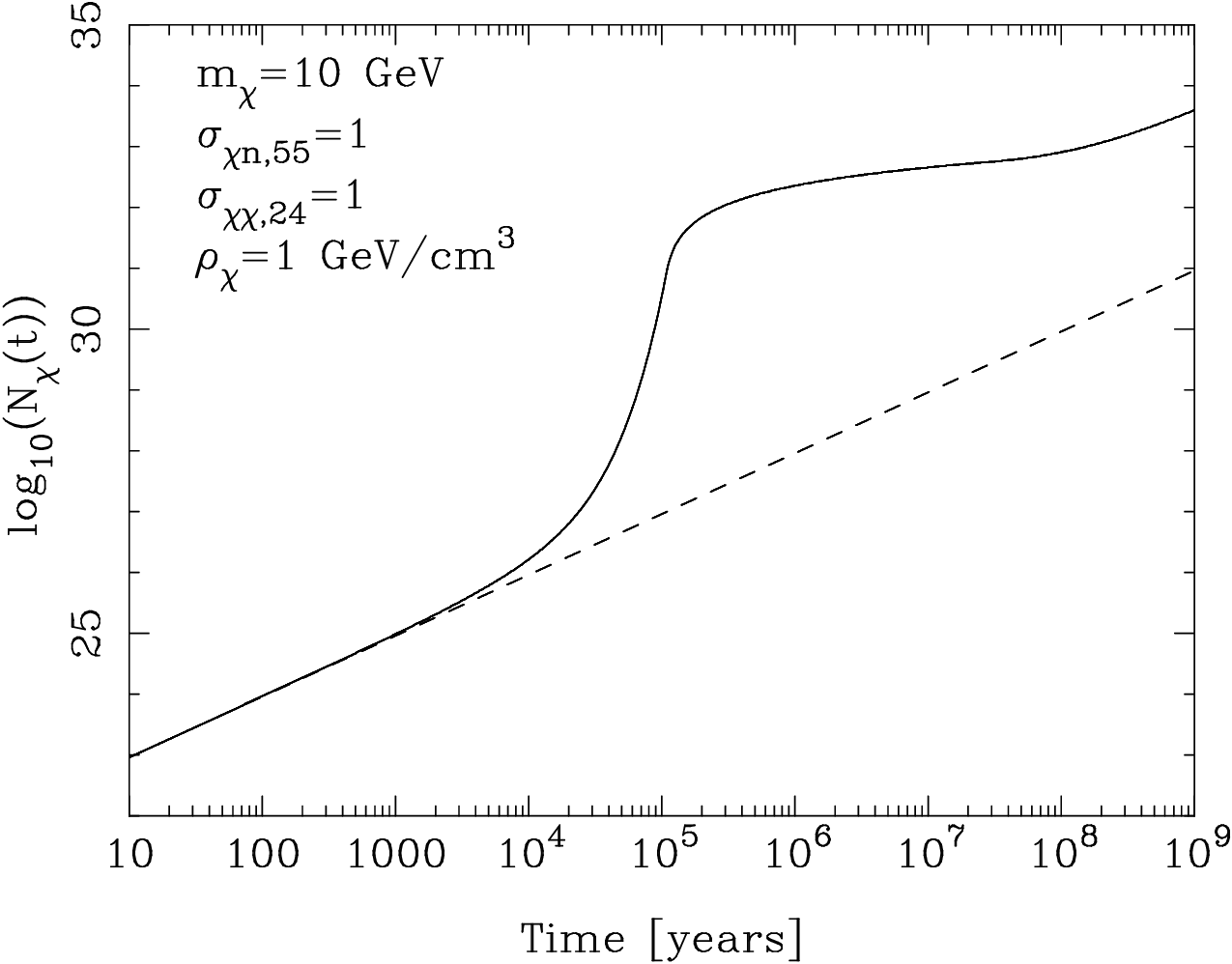}
   \includegraphics[angle=270,scale=0.38]{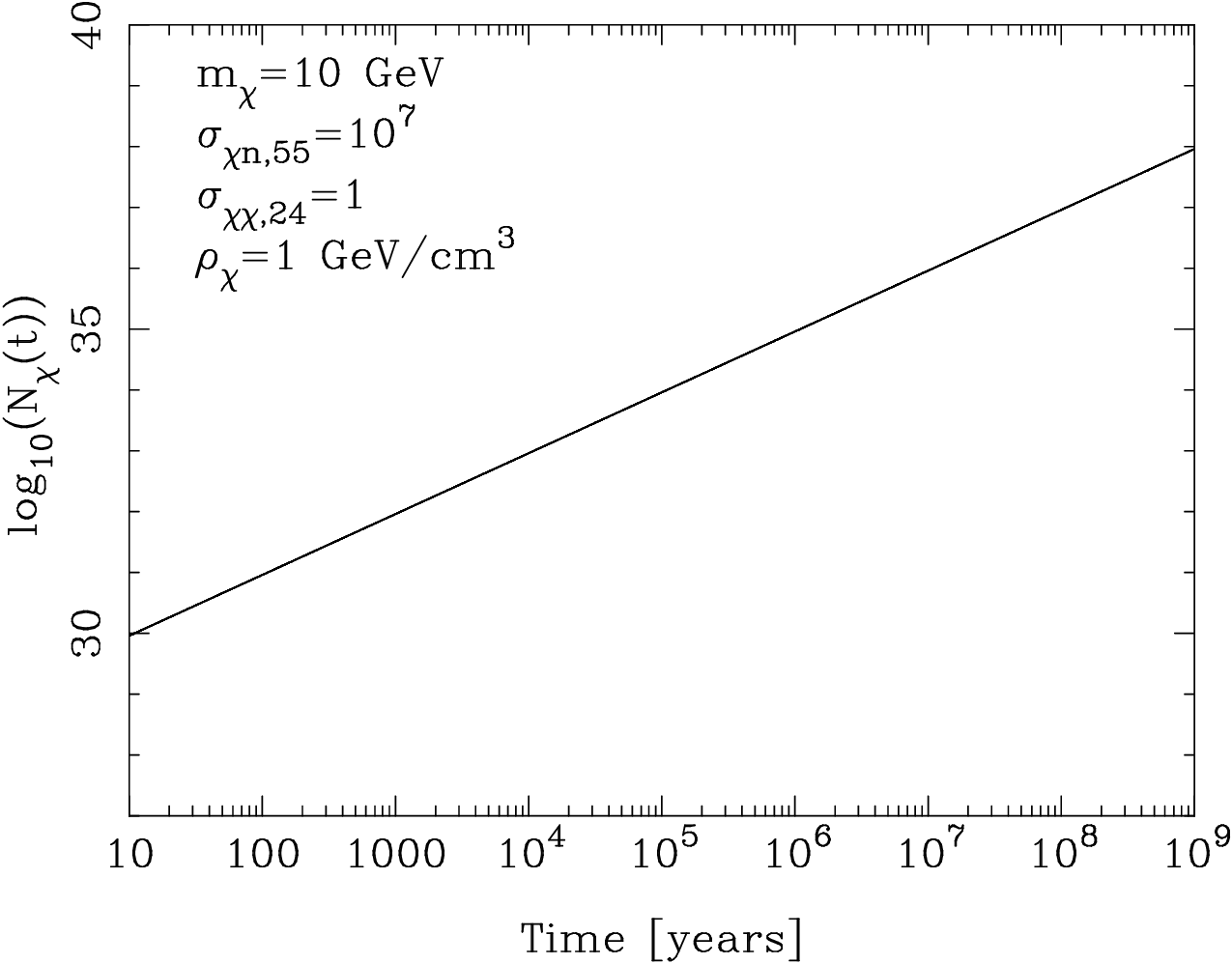}
   \caption{Same as Fig. 5, but for $m_\chi = 10$ GeV.} 
\label{f10gev}
\end{figure*}

\begin{figure*}
\centering
  \includegraphics[angle=270,scale=0.38]{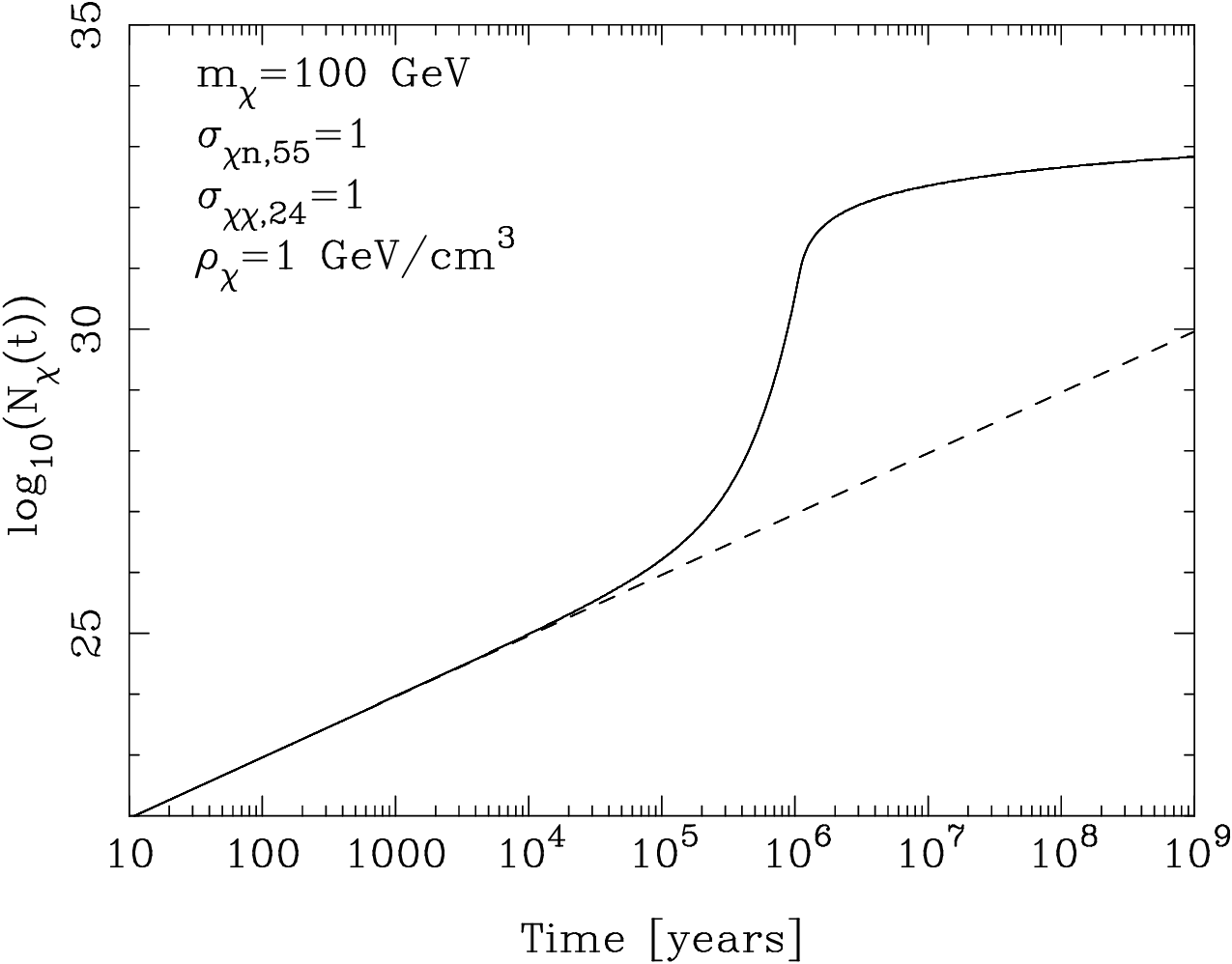}
   \includegraphics[angle=270,scale=0.38]{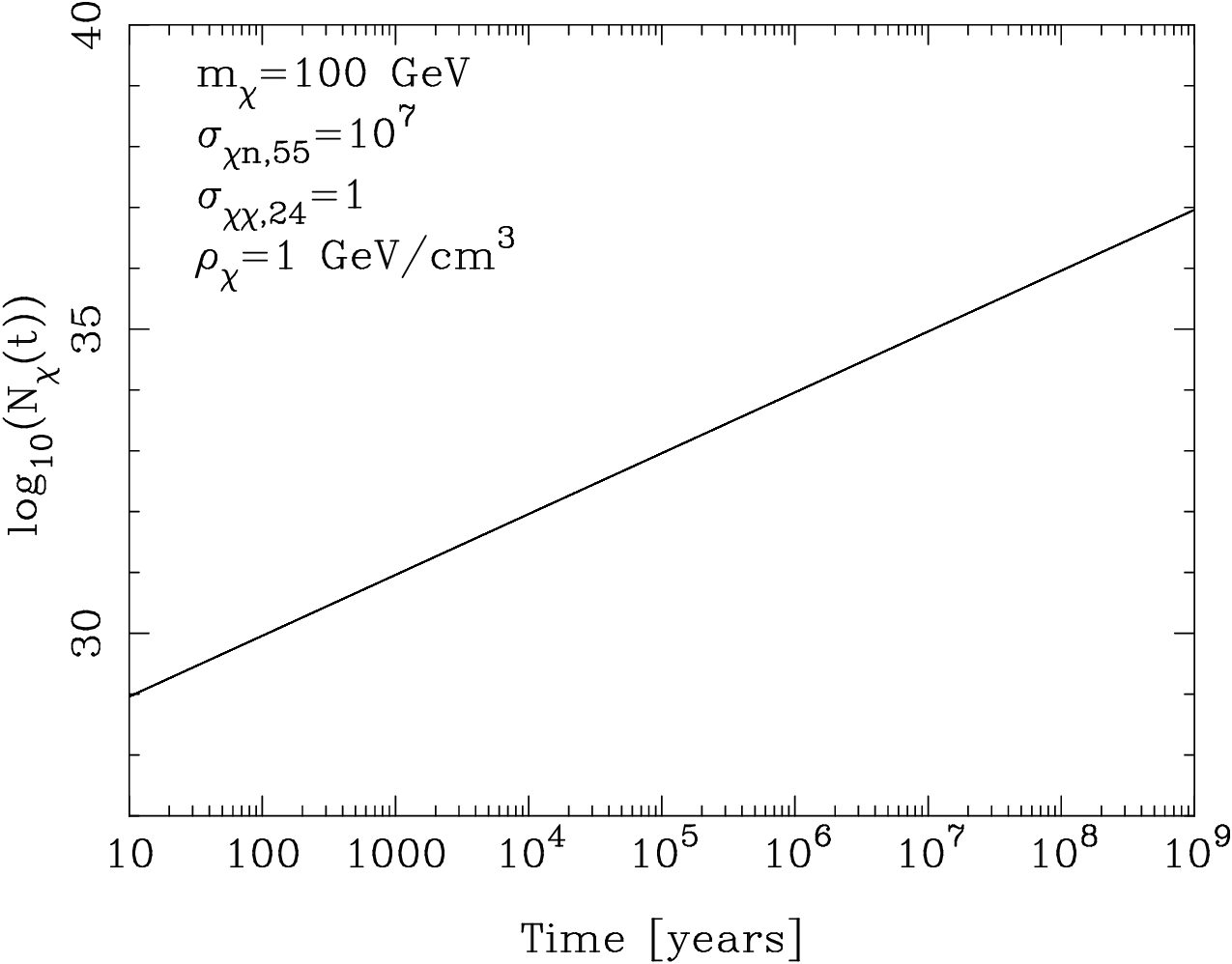}
   \caption{Same as Fig. 5, but for $m_\chi =100$ GeV.} 
\label{f100gev}
\end{figure*}

 Fig. \ref{xntg} shows log$_{10}(N_\chi(t_{max}))$ { and log$_{10}(N_\chi (t_G)$} as a function of $m_\chi$ for $\rho_\chi=10^3$ GeV/cm$^3$. 
In this figure, we have chosen 
$\sigma_{\chi n,55}=100$ and $\sigma_{\chi\chi,24}=10^{-3}$ and 1, but $N_\chi(t_{max})$ as a function of $m_\chi$
is nearly equal to the curve shown in Fig. \ref{xntg} for a wide range of $\sigma_{\chi n,55}$ and $\sigma_{\chi\chi,24}$.
Whether $t_G<t_{th}$ or $t_G>t_{th}$, for $t_{max}>>t_{th}$, 
we can approximate eq. (\ref{eq:Ntmax})  or eq. (\ref{eq:Ntmaxp}) by
\be
N_\chi(t_{max})\simeq\Bigl( C_c+C_s\frac{\pi r_{th}^2}{\sigma_{\chi\chi}}\Bigr) t_{max}\ .
\ee
When the term proportional to $C_s$ is most important, this simplifies to
\begin{eqnarray}
\label{eq:Ntmaxsimple}
N_\chi(t_{max})&\simeq & C_s\frac{\pi r_{th}^2}{\sigma_{\chi\chi}} t_{max}\\
\nonumber
&\simeq &
3.5\times 10^{38} \, \frac{\rm GeV^2}{m_\chi^2}\, \frac{\rho_\chi}{10^3\ {\rm GeV/cm}^3}\, \frac{t_{max}}{10^9\ {\rm yr}},
\ee
independent of $\sigma_{\chi\chi}$ and $\sigma_{\chi n}$. { As an example, for $m_\chi=10$ GeV and $\sigma_{\chi n,55}=1$, eq. (\ref{eq:Ntmaxsimple}) applies to $\chi\chi$ cross sections in the range of $\sigma_{\chi\chi}\sim 10^{-33}-10^{-24}$ cm$^2$.}
{ Even when self-capture dominates,} eq. (\ref{eq:Ntmaxsimple}) is modified when
$t_{th}$ or $t_G' $ is close to $ t_{max}$. In Fig. \ref{xntg}, this is where the curve for $N_\chi(t_{max})$ deviates from the power 
law $N_\chi (t_{max})\sim m_\chi^{-2}$.

For masses below $m_\chi \simeq 7.4 $ GeV, the minimal limit for black
hole production is $1.5\times 10^{38}\, {\rm GeV^2/}{m_\chi^2}$, while
for     larger    masses,     we     need    $N_\chi(t_{max})>     N_b
^{BEC}(r=r_{th})=2.7\times  10^{36}$ for  $T=10^5\ K$.  The limits  on
$\sigma_{\chi n}$ as  a function of mass come from  where the solid or
dashed lines  cross these  limits represented by  the dotted  curve in
Fig.  \ref{xntg}.  Except   for  the  cross  sections   for  $\chi  n$
interactions which have  $t_{th}$ nearly equal to  $t_{max}$, there is
no  potential  to constrain  dark  matter  masses above  approximately
$m_\chi\simeq  13$ GeV,  for  small $\chi  n$  interactions and  large
$\chi\chi$ interactions, even with $\rho_\chi=10^3$ GeV/cm$^3$.

{  Our  detailed  solutions  of  $N_\chi(t)$  for  $C_a=0$,  shown  in
  eqs.  (\ref{eq:Ntmax}-\ref{eq:Ntg},  \ref{eq:Ntmaxpp}, \ref{eq:Ntmaxp}),  rely  on  $r_\chi  (t)$ assuming  $\chi  n$
  scattering  is solely  responsible for  dark matter  cooling in  the
  neutron star. Our result in eq. (\ref{eq:Ntmaxsimple}), that $N_\chi(t_{max})$ depends
  only  on the  thermalization  radius, the  dark  matter density  and
  $t_{max}$  when  $C_s$  is  most important,  gives  support  to  our
  approximate approach.  In the  regime where $\chi\chi$ scattering is
  responsible for DM capture by the  neutron star, only a small energy
  loss by the incident DM is required for capture. However, in the two
  particle scattering, the net change in  energy is zero. For $\chi n$
  scattering,  the additional  neutron energy  is dissipated  via $nn$
  interactions  among the  $\sim  10^{57}$  neutrons. With  $\chi\chi$
  scattering, where, e.g., $N_\chi^{Ch}\simeq 10^{36}$ for $m_\chi=10$
  GeV, the  energy loss of  the incident  $\chi$ is transfered  to the
  target DM,  leading to  capture of  the incident  dark matter but  
  not to
  evaporation of the target dark matter.  A more detailed evaluation of the
  energetics of  $\chi\chi$ scattering  may result in some changes in 
  the time
  evolution of the radius of the  dark matter sphere, however, as long
  as $t_{max}\gg t_{th}$, those details  are not important. In view of
  our rather pessimistic conclusion that  only a limited range of dark
  matter  mass   is  potentially   constrained,  even   with  strongly
  interacting  bosons with  $\rho_\chi \sim  10^3$ GeV/cm$^3$,  a more
  detailed evaluation of $r_\chi(t)$ does not seem merited.  }

\section{Conclusions}

Our limiting results for $\rho_\chi=10^3$ GeV/cm$^3$ using $N_b^{Ch}$ and $N_b^{BEC}(r=r_{th})$
are shown in Fig. \ref{limits}. The upper region
is excluded when $\sigma_{\chi\chi}=0$, because of $\chi n$ scattering. Thermalization is crucial for the
dark matter particles to become self-gravitating, thus 
the region of parameter space 
 for which 
the thermalization does not occur cannot be excluded by 
considering the collapse of the neutron star.  The 
 region labeled ``no thermalization" shows the region of
parameters for which $t_{th}>10^9$ yr. 
The vertical line in Fig. \ref{limits} { comes
from the Bullet Cluster limit when $\sigma_{\chi\chi ,24}=1$
in eq. (3.10), namely
that $m_\chi>0.5$ GeV.}

The upper
exclusion region in Fig. \ref{limits} is consistant with the 
limits in Ref. \cite{McDermott:2011vn} for our choice of 
 $t_{max}$.  This limit takes into 
account 
Bose-Einstein condensate. A similar shape for the exclusion region
without the Bose-Einstein condensate restriction (for $m_\chi>7.4$ GeV) appears in Ref. \cite{Kouvaris:2011fk}. For
$m_\chi<m_n$, the inclusion of 
Pauli blocking changes the slope of the limiting exclusion region
for $\sigma_{\chi\chi,24}=0$.}

In Fig. \ref{limits}, the middle region shows the potential exclusion
range  of $\sigma_{\chi  n}$ when  $\sigma_{\chi\chi,24}=1$ { (labeled
``excluded $\sigma_{\chi\chi,24}=1$''). The almost vertical line near
$m_\chi\sim 10$ GeV, beyond which we have no constraints, comes from
using $N_{min} = N_b^{BEC}(r=r_{th})$, where the dotted line in 
Fig. 8 crosses the solid line.  

While Fig. \ref{limits} is only for one choice of $\sigma_{\chi\chi,24}$,
we found the limits on $\sigma_{\chi n,55}$ as a function of $m_\chi$ are 
applicable for a wide range of $\sigma_{\chi\chi,24}$.}
A range of  large $\chi\chi$ interaction cross sections gives
an exponential increase similar to  that seen in Figs. 5-7, increasing
$N_\chi$  by  several   orders  of  magnitude  over   the  case  where
$\sigma_{\chi\chi,24}=0$.    { For  $m_\chi   =   10$   GeV and
$\sigma_{\chi   n,55}=100$,  the resulting $N_{\chi}(t_{max})$}   is  the   same  for
$\sigma_{\chi \chi  ,24}=10^{-9}-1$, resulting in $N_\chi(t_{max})$ being 
  larger
than $N_b^{BEC}(r=r_{th})$ and thus corresponding to the 
 the exclusion region.  Since $C_c\propto  \rho_\chi \sigma_{\chi
  n}$, when  $\rho_\chi$ decreases,  the region  with a  potential for
exclusion of bosonic dark matter decreases.  If $\rho_\chi$ is smaller
than $10^3$ GeV/cm$^3$ by a factor of $\sim 2$, we do not have any new
limits  on   self-interacting  asymmetric  bosonic  dark   matter  for
$t_{max}=10^9$  yr,  even with  the  minimal  Chandrasekhar limit  for
bosons. { Indeed, by changing any combination of 
 parameters like neutron star temperature, 
 mass, radius, and age ($t_{max}$)  and
the dark matter velocity dispersion in a way that it would give 
 an overall factor of two reduction in $N_\chi$ (approximated by
eq. (4.2)) the limits we obtain would not apply.}

\begin{figure}
\centering
  \includegraphics[angle=270,scale=0.38]{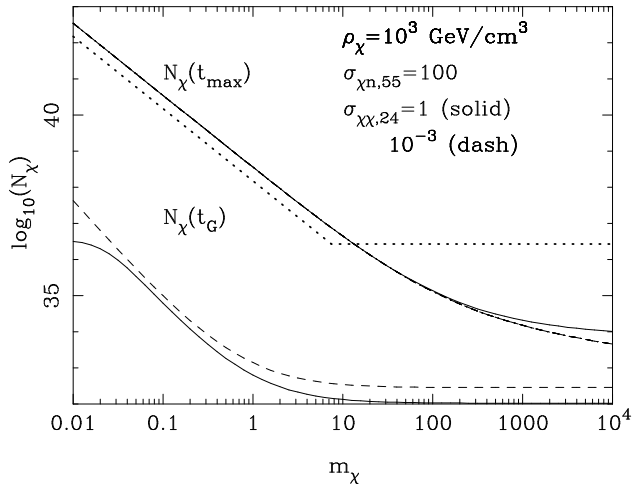}
   \caption{As a function of $m_\chi$, $\log_{10}(N_\chi(t_{max}))$ and $\log_{10}(N_\chi(t_{G}))$ for $\sigma_{\chi n,55}=100$, $\sigma_{\chi\chi,24}=
10^{-3}$ (dash)) and 1 (solid) for $\rho_\chi=10^3$ GeV/cm$^3$. The dotted line shows the limit above which
dark matter may collapse to a black hole within the neutron star.}
\label{xntg}
\end{figure}

\begin{figure}
\centering
  \includegraphics[angle=270,scale=0.38]{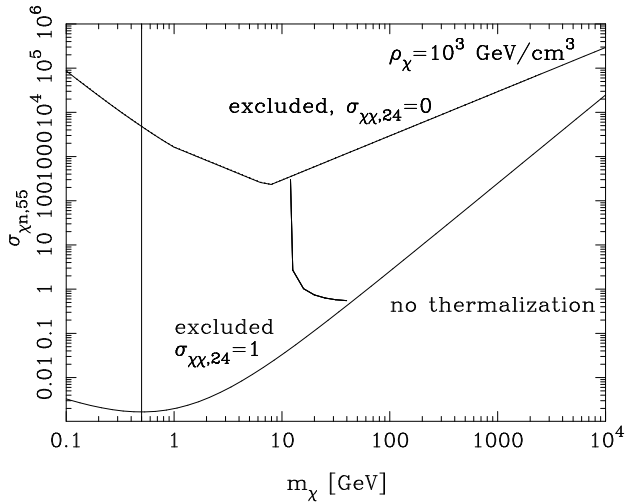}
   \caption{As a function of $m_\chi$, exclusion regions of $\sigma_{\chi n,55}$ for $\sigma_{\chi\chi,24}= 1$ and
$\rho_\chi=10^3$ GeV/cm$^3$, using  $N_b^{Ch}$ and $N_b^{BEC}(r=r_{th})$ for asymmetric bosonic dark matter. 
The region to the left of the vertical line { at $m_\chi=0.5$ GeV} is excluded by the Bullet Cluster limit { when
$\sigma_{\chi\chi,24}= 1$.} }
\label{limits}
\end{figure}

{ The minimal Chandrasekar bound for bosons 
 given by eq. (2.6)  does not apply in case of perturbative repulsive
$\lambda \phi^4$ models for $\chi\chi$ self interaction.  
A repulsive interaction can provide extra pressure and
increase the minimum number of dark matter particles required to form a black hole \cite{Mielke2000}. 
An example of such contraints appears in, for example, 
Ref. \cite{Kouvaris:2011fk} 
where the resulting $\chi\chi$
cross sections excluded by neutron stars are in the range of $\sigma_{\chi\chi,24}\lsim 10^{-40}$.
Clearly with the large, nonperturbative cross sections that we 
considered, the limits are different 
from a perturbative $\lambda \phi^4$ model considered in 
Ref. \cite{Kouvaris:2011fk}.  

{ Evaporation by a newly formed black hold by Hawking radiation will dominate accretion unless
$N_\chi>N_{BH}^{crit}=5.7\times 10^{36}$ GeV/$m_\chi$ \cite{Kouvaris:2011fk}. We note that  
$N_b^{Ch}>N_{BH}^{crit}$ for $m_\chi<26$ GeV. The more restrictive self-gravitation limit for BEC, $N_{min}=N_b^{BEC}(r=r_{th})=
2.7\times 10^{36}$ is applied when 7.4 GeV$<m_\chi<1.26\times
10^4$ GeV (the dotted line in Fig. 8). Therefore, for the whole exclusion region in Fig. 9, the Hawking radiation limit 
$N_\chi(t_{max})>N_{BH}^{crit}$ is satisfied.}

Dark matter  self-interactions can  make significant  contributions to
the accumulation of dark matter in a neutron star when the dark matter
density  is  large, $\rho  \sim  10^3$  GeV.   We have  evaluated  the
accumulation including the consequences of a time dependent radius for
the  dark matter  distribution in  the neutron  star.  {  The quantity
  $r_\chi(t)$ in eq. (\ref{eq:rchit1}) follows from approximating the cooling of DM
  as coming from only dark  matter-nucleon scattering. As discussed in
  Sec. IIB,  only small energy  losses of  DM incident on  the neutron
  star and cloud of captured dark matter are required for capture. The
  net  change in  energy in  $\chi\chi$ scattering  is zero;  the dark
  matter eventually thermalizes with neutrons.

Our main result  for $t_{max}\gg t_{th}$, when $C_c$  can be neglected
relative to  the self-capture  term, is  $N_\chi (t_{max})  \simeq C_s
(\pi   r_{th}^2/c_{\chi\chi})t_{max}$  (eq.   (\ref{eq:Ntmaxsimple})),  independent   of
$\sigma_{\chi\chi}$.   A   more  detailed   evaluation  of   the  time
dependence of $r_\chi(t)$ including the  effect of the energy transfer
to  the already  captured DM  particles is  beyond the  scope of  this
paper, however, the form of eq. (\ref{eq:Ntmaxsimple}) for $N_\chi(t_{max})$ points to a
conclusion that a refinement of  $r_\chi(t)$ is unlikely to weaken our
limits when $t_{max}\gg t_{th}$.}

{ For  $\rho_\chi=10^3$ GeV/cm$^3$,}  we have found  a new  portion of
parameter   space    that   is    excluded   using    $N_b^{Ch}$   and
$N_b^{BEC}(r=r_{th})$,  for   example,  the  region  where   the  dark
matter-nucleon interaction cross section  is between 10$^{-52}$ cm$^2$
to 10$^{-57}$ cm$^2$ and dark matter self-interaction cross section is
greater  than   $\sim$10$^{-33}$  cm$^2$  for  $m_\chi=10$   GeV.   As
discussed earlier,  this region is interesting  because thermalization
takes a  long enough  time that  self-interactions play  a significant
role in capturing dark matter particles.

Appreciable   dark  matter   accumulation  through   self-interactions
requires large  dark matter  densities.  For  stars in  the core  of a
globular cluster, the  dark matter density could be as  high as $10^3$
GeV/cm$^3$ \cite{McCullough}. If the dark  matter density is this high
and  the minimal  limits of  $N_b^{Ch}$ and  $N_b^{BEC}(r=r_{th})$ are
applicable,   the   dark   matter  self-interaction   cross   section,
$\sigma_{\chi\chi}$, is  constrained when it  is in the  range between
$10^{-33}$ cm$^2$  and $10^{-24}$ cm$^2$,  which is several  orders of
magnitude more restrictive than the Bullet Cluster limit.
 
Finally, as  we have noted,  we have used eq.  (\ref{chandra_bos}) and
$N_b^{BEC}(r=r_{th})$  to constrain  bosonic  dark  matter in  neutron
stars.  Recent work  has considered  specific models  for dark  matter
self-interactions that  rely on  perturbation theory, and  a parameter
space of  repulsive couplings is ruled  out by the observation  of old
very   old  neutron   stars  \cite{Bramante,review,   bell}.  A   more
sophisticated   Chandrasekhar   limit   for   bosons   would   include
modifications  in   the  case  of  strongly   interacting  bosons,  or
attractive        interactions,       for        specific       models
\cite{review,csw,Mielke2000}.   We have  focused on  the Chandrasekhar
limit with relativistic bosons as the  starting point for the study of
self-interacting dark  matter.  Only  a limited  range of  dark matter
mass is potentially constrained with this approach, even with strongly
interacting  bosons and  only for  $\rho_\chi\sim 10^3$  GeV/cm$^3$ or
larger dark  matter densities.  An additional  interesting possibility
of detecting  asymmetric dark  matter in the  minimal self-interacting
dark   matter   model   with   gamma-rays   has   been   proposed   in
Ref. \cite{kusenko}.

\begin{acknowledgments}
We would like to thank  Dimitrios Psaltis, Dennis Zaritsky and Kathryn
Zurek for useful discussions.  IS would like to thank the Aspen Center
for Physics,  where part of  this work took place.   T.G. acknowledges
support from Bilim  Akademisi - The Science Academy,  Turkey under the
BAGEP program.  This research was supported by US Department of Energy
contracts  DE-FG02-91ER40664,   DE-FG02-04ER41319,  DE-FG02-04ER41298,
DE-FG03-91ER40662, DE-FG02-13ER41976,  and DE-SC0010114.
\end{acknowledgments}


\end{document}